\documentclass[10pt,twocolumn,journal]{IEEEtran}
 \usepackage{setspace}

\usepackage{footnote}
\usepackage{bbm}
\usepackage{amsmath,graphics,amssymb,epsfig,subfigure,color,amsthm,cite}
\usepackage{array}
\usepackage{multirow}
\usepackage{lipsum}
\usepackage{mathtools}
\usepackage{cuted}
\usepackage{enumerate}
\usepackage[ruled]{algorithm2e}
\SetKwInOut{Input}{Input}
\SetKwInOut{Output}{Output}
\DontPrintSemicolon

\usepackage{bm}
\usepackage[cmintegrals]{newtxmath}
\usepackage{tablefootnote}

\newcommand{\rom}[1]{\uppercase\expandafter{\romannumeral #1 \relax}}

\newtheorem{thm}{Theorem}

\begin{document}
\title{Distributed Precoding Using Local CSIT for MU-MIMO Heterogeneous Cellular Networks   } 

\author{Deokhwan Han, ~\IEEEmembership{Student Member,~IEEE,}  and Namyoon Lee,~\IEEEmembership{Senior Member,~IEEE,} \\
  \thanks{D. Han and N. Lee are with the Department of Electrical Engineering, POSTECH, Pohang, Gyeongbuk 37673, South Korea  (e-mail: \{dhhan, nylee\}@postech.ac.kr).}
	}

\maketitle
\begin{abstract}
Cell densification is a key driver to increase area spectral efficiencies in multi-antenna cellular systems. As increasing the densities of base stations (BSs) and users that share the same spectrum, however, both inter-user-interference (IUI) and inter-cell interference (ICI) problems give rise to a significant loss in spectral efficiencies in such systems. To resolve this problem under the constraint of local channel state information per BS, in this paper, we present a novel noncooperative multi-user multiple-input multiple-output (MIMO) precoding technique, called \textit{signal-to-interference-pulse-leakage-pulse-noise-ratio (SILNR)} maximization precoding.  The key innovation of our distributed precoding method is to maximize the product of SILNRs of users per cell using local channel state information at the transmitter (CSIT). We show that our precoding technique only using local CSIT can asymptotically achieve the multi-cell cooperative bound attained by cooperative precoding using global CSIT in some cases. We also present a precoding algorithm that is robust to CSIT errors in multi-cell scenarios. By multi-cell system-level simulations, we demonstrate that our distributed precoding technique outperforms all existing noncooperative precoding methods considerably and can also achieve the multi-cell bound very tightly even with not-so-many antennas at BSs.

 \end{abstract}

\IEEEpeerreviewmaketitle
\section{Introduction}

  Heterogeneous cellular networks (HetNets) are promising solutions for achieving high data rates ubiquitously\cite{ghosh2012heterogeneous, dhillon2013downlink,andrews2014overview,hosseini2013massive,chen2017scalable}. HetNets are comprised of distinct network tiers, each with different transmission power and number of antennas. By deploying low power base stations (BSs) (e.g., pico and femto BSs) overlaid onto macrocells, HetNets considerably increase area spectral efficiencies by cell-splitting gains. When operating these small BSs using the same frequency/time resources with macro-cells, significant intra-tier and inter-tier interference problems take place. This interference problem makes the area spectral efficiency gains dwindle. As a result, an effective interference mitigation technique is indispensable to obtain the area spectral efficiency gains in HetNets \cite{adhikary2014spatial ,lopez2011enhanced,gupta2014downlink}.

  Massive multiple-input multiple-output (MIMO) is an effective solution to resolve this interference problem by exploiting a large degree of freedom in the spatial domain \cite{adhikary2013joint,larsson2014massive,rusek2013scaling}. With time-division-duplexing (TDD) operation, it has shown in \cite{marzetta2010noncooperative} that simple precoding techniques with perfect channel state information at the transmitter (CSIT) are sufficient to eliminate both inter-user-interference (IUI) and inter-cell-interference (ICI) when using an infinite number of antennas. This result implies that noncooperative precoding with perfect CSIT of the associated users in a cell can be asymptotically optimal when the number of antennas is sufficiently larger than that of active downlink users per cell \cite{hoydis2013massive}. In HetNets, however, the number of antennas of BSs in small-cells cannot be many by the cost and hardware limitations \cite{ghosh2012heterogeneous, dhillon2013downlink,andrews2014overview}. Besides, the density of active users in the hotspots is relatively high. In this case, the channel hardening effects are not sufficiently pronounced except for the ray-based channel model cases \cite{li2019massive}. In general, the simple precoding methods fail to successfully eliminate both ICI and IUI in massive MIMO networks \cite{marzetta2010noncooperative,hoydis2013massive}.
  
  
     This paper focuses on a multi-user MIMO (MU-MIMO) HetNet where the macro-cell tier is overlaid with small-cells, each small-cell BS is equipped with a few antennas. The user density of the small-cell areas is higher than that of the other areas. With a limited number of antennas at the BSs, most prior works have focused on cooperative precoding strategies to eliminate both IUI and ICI, which provides considerable gains in the sum-spectral efficiency. In practice, however, these cooperative precoding strategies are undesirable when considering the overheads for the CSI exchange among BSs via backhaul signaling; the gains by the cooperation can disappear \cite{lee2014spectral,park2015cooperative}. Instead, the precoding strategies using local CSIT have significant merit to simultaneously reduce IUI and interference leakage to other cells' users without causing any signaling overheads. 
     
     One of the challenges in designing noncooperative precoding strategies is that the ICI leakage mitigation using local CSIT does not necessarily maximize the sum-spectral efficiency; thereby, the performance gap between the cooperative and noncooperative methods can be consequential. Therefore, finding a noncooperative precoding strategy that can closely attain the performance gain of the cooperative precoding is a significant yet challenging problem. This paper tackles this problem and shows finding such a noncooperative precoding strategy is affirmative in some instances.


     

%
  
\subsection{Prior Works}
There are extensive prior studies for multi-cell linear precoding methods using local CSIT. The simplest method is maximum ratio transmission (MRT) \cite{lo1999maximum}, which is also known as matched filtering (MF) precoding. To employ MRT, each BS only requires to have the local CSIT of its cell. In a multi-cell massive MIMO setting, where the number of BS antennas is much larger than that of users, it has shown in \cite{marzetta2010noncooperative} that this simple precoding can asymptotically eliminate both IUI and ICI under perfect local CSIT assumption.  Thanks to its simple precoding structure, the analytical expressions for achievable rates have been derived in closed-forms as a function of relevant system parameters in massive MIMO settings \cite{bjornson2014designing,lim2015performance}.


Zero-forcing (ZF) \cite{spencer2004zero} is another popular precoding method to eliminate IUI using local CSIT. Unlike MRT precoding, it can entirely remove IUI regardless of the number of BS antennas by selecting the number of users that is not larger than that of BS. Particularly, when the number of users is sufficiently larger than that of BS antennas, ZF precoding with semi-orthogonal user selection \cite{yoo2006optimality} has shown to asymptotically achieve the optimal capacity scaling law attained by \cite{costa1983writing} (DPC). In a massive MIMO setup, in which a BS has an infinite number of antennas, ZF precoding can maximize the sum-spectral efficiency under the perfect CSIT assumption \cite{marzetta2010noncooperative}. When the BS has a not-so-large number of antennas compared to the number of users such as HetNets \cite{ghosh2012heterogeneous, dhillon2013downlink,bogale2016massive}, ZF precoding is not effective to mitigate both IUI and ICI simultaneously because of the inefficient utilization of the spatial degrees of freedom.



Signal-to-leakage-plus-noise-ratio (SLNR) precoding \cite{sadek2007leakage} is an effective method to suppress both IUI and ICI using local CSIT in multi-cell MIMO networks. It turns out that this precoding is equivalent to the minimum mean square error (MMSE) precoding (or regularized ZF precoding) under uniform power allocation. In a multi-cell massive MIMO setting, the multi-cell MMSE precoding method has been proposed in \cite{jose2011pilot} by taking into account pilot contamination effects. Particularly, in cell-free massive MIMO setting \cite{bjornson2019new}, SLNR precoding is an attractive precoding method using local CSIT because of its scalability and ICI mitigation capability. The major limitation is that SLNR maximization precoding does not necessarily maximize sum-spectral efficiency.  


Several linear precoding algorithms have been proposed to maximize the sum-spectral efficiency in single-cell and multi-cell multi-user MIMO systems \cite{christensen2008weighted,jose2011pilot,choi2018joint,choi2019joint}. Finding the global optimal linear precoder that maximizes the sum-spectral efficiency is NP-hard. The weighted-MMSE (WMMSE) precoding is the most popular sub-optimal precoding technique in the sum-spectral efficiency maximization problem \cite{christensen2008weighted}. Thanks to the equivalence between the sum-spectral efficiency maximization problem and the WMMSE minimization, an alternating minimization algorithm has been proposed, which converges to a local-optimal solution. However, this precoding cannot be applicable when the number of antennas is massive at the BS because it requires very high computational complexity. Recently, inspired by principal component analysis (PCA), a novel low-complexity algorithm called generalized power iteration precoding (GPIP) has been presented, which guarantees the first-order optimality for the sum-spectral efficiency maximization problem under perfect and noisy CSIT \cite{choi2018joint}. This precoding method has also been extended to a multi-cell scenario with pilot contamination effects  \cite{choi2019joint}. This multi-cell precoding method, however, requires to exchange the CSI among BSs, which causes a significant signaling overhead.

%
%

\subsection{Contributions}

The main contributions of this paper is summarized as follows:
 
\begin{itemize}

\item We introduce a new performance metric to effectively mitigate both IUI and ICI in a distributed manner using local CSIT for MU-MIMO HetNets. The new performance metric is the ratio of signal-to-interference-pulse-leakage-pulse-noise-ratio (SILNR). Intuitively, SILNR measures the ratio between the desired signal power and the sum of IUI, the interference leakage, and the noise power. However, this new metric significantly differs from the conventional SLNR  \cite{sadek2007leakage,jose2011pilot} in two aspects. First, while the IUI in the definition of the existing SLNR is treated as the intra-cell leakage interference (i.e., the uplink IUI), the IUI in the SILNR is the exact downlink IUI. Second, the interference leakage term in the SLNR is the sum of interference leakage signals to the individual users in the other cells. In contrast, the interference leakage term in our SILNR is the geometric mean of the interference leakage signals to the other cell users. We show that maximizing the product of SILNRs of users per cell using local CSIT can achieve the identical sum spectral efficiency with the cooperative precoding method using global CSIT in a two-cell MU-MIMO system under certain cases.

	\item We generalize this precoding method for a multi-cell setting under noisy CSIT assumption. Unfortunately, the product of SILNRs maximization is a non-convex (and even NP-hard) optimization problem similar to the sum-spectral efficiency maximization problem. To design the precoding method, we first derive the first- and second-order necessary conditions for this non-convex optimization problem. Using the derived conditions, we present a low-complexity iterative algorithm that guarantees to converge a locally-optimal solution. The key innovation of the proposed precoding is to identify the joint solutions for the scheduled users per cell,  the beamforming vectors of them, and the power allocated to each beam using local and noisy CSIT.

	\item Using both link-level and system-level simulations, we exhibit that the proposed precoding method significantly outperforms the existing distributed precoding techniques, including MRT, ZF, SLNR, and the sum-spectral efficiency maximization precoding per cell in both perfect and imperfect CSIT scenarios. One remarkable observation is that the proposed method asymptotically achieves the upper bound performance attained by the multi-cell cooperative precoding method when increasing the number of macro BS antennas. This result confirms that our noncooperative precoding technique using local CSIT can be a pragmatic solution to resolving dense cellular networks' interference problems.  
 	
\end{itemize}
\subsection{Notations}
 
Throughout this paper, we use the following notations. The $\mathbb{C}$ denotes a set of a complex number; $\mathbb{R}$ denotes the set of a real number. The $\otimes$ is the Kronecker product operation. We use $\mathbb{E}_{{\bf x}}\left[{\bf x}\right]$ to denote the expectation of a random vector ${\bf x}$. The ${\sf Re}\left\{ x \right\}$ means the real part of a complex scalar $x$. In addition, we use $\lambda_{\sf min}\left({\bf A}\right)$ and $\lambda_{\sf max}\left({\bf A}\right)$ to denote the minimum and maximum eigenvalue of matrix ${\bf A}$, respectively. A matrix ${\bf I}_N$ denote an $N\times N$ identity matrix. Also, ${\bf x}\sim\mathcal{CN\left({\bf m},{\bf R}\right)}$ indicates that the random vector ${\bf x}$ is distributed by complex Gaussian distribution with mean vector ${\bf m}$ and covariance matrix ${\bf R}$.

\section{System Model}
\begin{figure}[t]
\centering
\includegraphics[width=\linewidth]{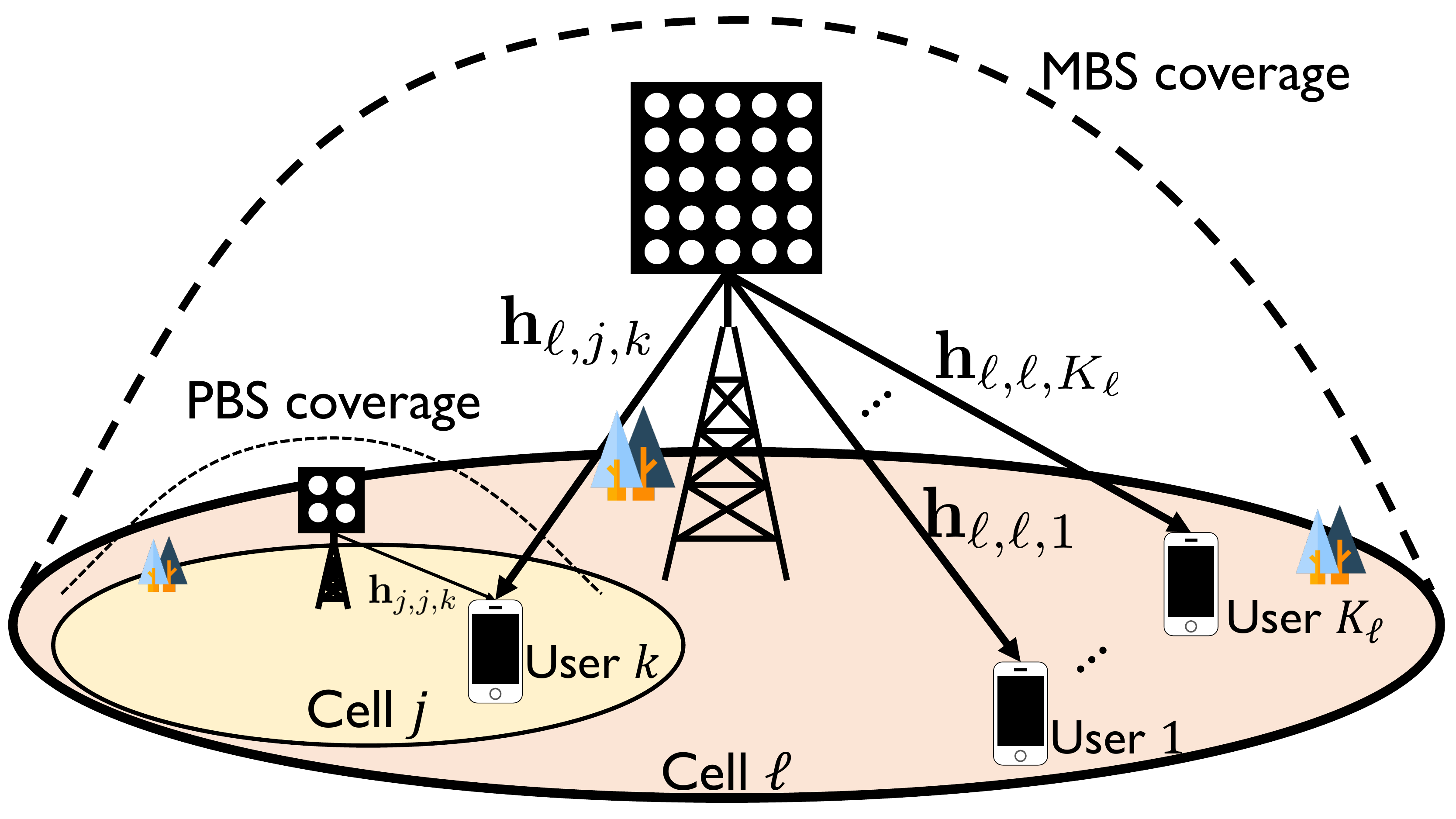}
\caption{An illustration of MU-MIMO HetNets.}
\label{System_Model}
\end{figure}


We consider a heterogeneous cellular network, which comprises of $L$ BSs each equipped with $N_{\ell}$ antennas for ${\ell}\in\mathcal{L} \triangleq\{1,\hdots,L\}$. The $\ell$th BS serves $K_{\ell}$ users with single antenna.

\subsection{Downlink Channel Model}
As illustrated in Fig. \ref{System_Model}, we denote the downlink channel vector from the $\ell$th BS to the $k$th user in the $j$th cell by ${\bf h}_{\ell,j,k}\in\mathbb{C}^{N_{\ell}\times 1}$. This downlink channel is modeled as \begin{align}
    {\bf h}_{\ell,j,k} = \beta_{\ell,j,k}^{\frac{1}{2}}{\bf g}_{\ell,j,k},
\end{align}
where $\beta_{\ell,j,k}\in \mathbb{R}$ is a large scale fading coefficient and ${\bf g}_{\ell,j,k}\in\mathbb{C}^{N_{\ell}\times 1}$ is a small scale fading, which is distributed as ${\bf h}_{\ell,j,k}\sim\mathcal{CN}(0,\beta_{\ell,j,k}{\bf R}_{\ell,j,k})$ with ${\bf R}_{\ell,j,k} = \mathbb{E}\left[{\bf g}_{\ell,j,k}{\bf g}_{\ell,j,k}^{\sf H}\right]$  for $ \forall \ell,j\!\!\in\!\!\mathcal{L}$ and $\forall k\in\mathcal{K}_{j}\triangleq\{1,\hdots,K_{j}\}$. This matrix captures the spatial correlation information on the channel. Under a stationary process assumption, it is typically obtained by using both angle-of-arrival (AoA) vectors of multipaths and the corresponding angular autocorrelation function.


\subsection{Local and Noisy CSIT Acquisition}

We present a process of acquiring local CSIT at each BS. For ease of exposition, we focus on the $\ell$th BS.  We define a subset $  \mathcal{C}_{\ell} \subset \mathcal{L}$ as the collection of BSs that use mutually orthogonal uplink pilot sequences with those of the $\ell$th BS for uplink channel training. This BS subset is assumed to be predetermined with a proper cell planning method; thereby, no dynamic BS cooperation is required.  Under the premise that $\ell$th BS has full knowledge of the orthogonal pilots, it can estimate the channel from the users associated with the $j$th BS where $j\in \mathcal{C}_{\ell}$. For example, let assume that the uplink users served by BS 1, BS 2, BS 3, and BS 4 use mutually orthogonal pilot sequences, then $\mathcal{C}_{1}=\mathcal{C}_{2}=\mathcal{C}_{3}=\mathcal{C}_{4}=\{1,2,3,4\}$. To guarantee the orthogonality, the uplink pilot length satisfies the condition of $\tau_{\rm u}\geq \sum_{j=1}^4K_{j}$. Then, BS $\ell$ can estimate the channels from the users associated with BS $j$ where $j\in \{1,2,3,4\}$. Thanks to the channel reciprocity in TDD mode, each BS obtains the downlink channel vectors from the uplink channel estimates with a proper RF circuit calibration process.



%

Applying the MMSE estimator, the estimated downlink channel vector from the $\ell$th BS to the $k$th user in the cell is
\begin{align}
    {\bf \hat h}_{\ell,\ell,k} &= {\bf h}_{\ell,\ell,k} - {\bf e}_{\ell,\ell,k},
\end{align}
where ${\bf e}_{\ell,\ell,k}$ is the estimation error vector and it is distributed as complex Gaussian with zero-mean and covariance matrix ${\bf \Phi}_{\ell,\ell, k}={\mathbb{E}\left[{\bf e}_{\ell,\ell,k}{\bf e}_{\ell,\ell,k}^{{\sf H}}\right]}\!\in\! \mathbb{C}^{N_{\ell}\times N_{\ell}}$, i.e., $\mathcal{CN}\left({\bf 0}, {\bf \Phi}_{\ell,\ell,k}\right)$.  The error covariance matrix is \cite{yin2013coordinated,hoydis2011massive}:
\begin{align}
   &{\bf \Phi}_{\ell,\ell,k} = {\beta}_{\ell,\ell, k}{\bf R}_{\ell,\ell,k}\nonumber\\ &-{\beta}_{\ell,\ell,k}^2{\bf R}_{\ell,\ell,k}
    \left(\sum_{ (j,i) \in {\bar{\mathcal{C}}}_{\ell} }{\beta}_{\ell,j,i}{\bf R}_{\ell,j,i}+\frac{\sigma^2}{\tau_{\rm u}p_{\sf ul}}{\bf I}_N
    \right)^{-1}{\bf R}_{\ell,\ell,k},\label{eq:Est_Err_Cov}
\end{align}
where $p_{\sf ul}$ is the pilot transmission power and ${\bar{\mathcal{C}}}_{\ell}$ is the collection of users who use the non-orthogonal uplink pilots with that of the $k$th user in the $\ell$th cell, i.e., ${\bar{\mathcal{C}}}_{\ell}=\left\{(j,i)~|~\forall j\in\mathcal{L}\setminus\mathcal{C}_{\ell}, \forall i\in\mathcal{K}_{j}\right\}$.

\subsection{Ergodic Downlink Spectral Efficiency with Noisy CSIT}
Suppose $\tau_{\rm c}$ be a channel coherence time interval. We also let ${\bf x}_{\ell}[t] \in \mathbb{C}^{N_{\ell} \times 1}$ be the transmit signal of the $\ell$th BS using the $t$th time slot where $t\in\left[1, \tau_{\rm c}\right]$.  The $\ell$th BS transmits $K_{\ell}$ independent data symbols $\left\{s_{\ell,1}[t],\ldots,s_{\ell,K_{\ell}}[t]\right\}$ using time slot $t$ along with precoding vectors $\{ {\bf f}_{\ell,1},\ldots, {\bf f}_{\ell,K_{\ell}}\}$. The precoding vectors are constructed using noisy and local CSIT at the $\ell$th BS, i.e., $\left\{ {\bf \hat h}_{\ell,\ell,1},{\bf \hat h}_{\ell,\ell,2},\ldots, {\bf \hat h}_{\ell,\ell,K_{\ell}} \right\}$ and $\left\{ {\bf   \hat  h}_{\ell,j,i}\right\}$ where $j\in \mathcal{C}_{\ell}$ and $i\in \mathcal{K}_{j}$.  We assume that each data symbol $s_{\ell,k}[t]$ is drawn from a Gaussian codebook with transmit power $P_{\ell}$, i.e., $s_{\ell,k}[t]\sim \mathcal{CN}(0, P_{\ell})$. In addition, the linear precoding vectors satisfy the condition of $\sum_{k=1}^{K_{\ell}}\|{\bf f}_{\ell,k}\|_2^2 \leq1$  to meet the power constraint per BS. Then, the transmit signal of the $\ell$th BS at the $t$th time slot is
\begin{align}
    {\bf x}_{\ell}[t] = \sum_{k=1}^{K_{\ell}}{\bf f}_{\ell,k}s_{\ell,k}[t].
\end{align}
Then, the received signal of the $k$th user in the $\ell$th cell is given by
\begin{align}
    y_{\ell,k}[t]\! &=\! \sum_{j=1}^{L}{\bf h}_{j,\ell,k}^{\sf H}{\bf x}_j[t] +
    n_{\ell,k}[t], 
\end{align}
where $n_{\ell,k}[t]$ is the additive complex Gaussian noise with zero-mean and variance $\sigma^2$, i.e., $\mathcal{CN}\left(0,\sigma^2\right)$. Then, the signal-to-interference-plus-noise ratio (SINR) of the $k$th user in the $\ell$th cell is given by
\begin{align}
    {\sf SINR}_{\ell,k} = \frac{\left|{\bf h}_{\ell,\ell,k}^{\sf H}{\bf f}_{\ell,k}\right|^2}{\sum_{i\ne k}^{K_{\ell}}\left|{\bf h}_{\ell,\ell,k}^{\sf H}{\bf f}_{\ell,i}\right|^2+\sum_{j\ne \ell}^{L}\sum_{i=1}^{K_j}\frac{P_j}{P_{\ell}}\left|{\bf h}_{j,\ell,k}^{\sf H}{\bf f}_{j,i}\right|^2+\frac{\sigma^2}{P_{\ell}}}. \label{eq:SINR}
\end{align}
Let $\tau_{\rm d}$ be the downlink channel training length. Using the orthogonal downlink channel training sequence with length $\tau_{\rm d}\geq \sum_{j \in \mathcal{C}_{\ell}} K_j$, i.e., demodulation reference signals in Long-Term Evolution (LTE) systems, the $k$th downlink user associated with the $\ell$th BS can estimate the precoded downlink channel state information at receiver (CSIR), i.e., ${\bf h}_{\ell,\ell,k}^{\sf H}{\bf f}_{\ell,k}$.  For simplicity, we assume that each downlink user has perfect CSIR for the precoded channel for the ease of exposition. To incorporate the effect of imperfect CSIR, one can use the notion of generalized mutual information introduced in \cite{medard2000effect,yoo2004mimo} to redefine SINR per user with the channel estimation error variance.

With noisy and local CSIT at the $\ell$th BS, i.e., ${\hat{\mathcal{H}}}_{\ell}=\left\{ {\bf\hat  h}_{\ell,j,i}~|~j\in\mathcal{C}_{\ell}, i\in\mathcal{K}_{j}\right\}$ corresponding to the ${{\mathcal{H}}}_{\ell}=\left\{ {\bf  h}_{\ell,j,i}~|~j\in\mathcal{C}_{\ell}, i\in\mathcal{K}_{j}\right\}$, it can  estimate the {instantaneous} spectral efficiency of the $k$th downlink user \cite{choi2018joint,choi2019joint}:
\begin{align}
	R_{\ell,k}\left({\hat{\mathcal{H}}}_{\ell}\right) = \mathbb{E}_{ {{{\mathcal{H}}}_{\ell} | {\hat{\mathcal{H}}}_{\ell} } } \left[\log_2\left(1+{\rm SINR}_{\ell,k}\right)\mid {\hat{\mathcal{H}}}_{\ell} \right], \label{eq:avg_SE}
\end{align}
where the expectation is taken over the CSIT error distribution, i.e., ${\bf e}_{\ell,\ell,k}\sim \mathcal{CN}\left({\bf 0}, {\bf \Phi}_{\ell,\ell,k}\right)$ and ${\bf e}_{\ell,j,i}\sim \mathcal{CN}\left({\bf 0}, {\bf \Phi}_{\ell,j,i}\right)$.  This quantity measures the average spectral efficiency over the CSIT error distribution for a given estimates of CSIT.  Therefore, by taking the expectations over every fading state, the effective ergodic spectral efficiency is given by
 \begin{align}
{\bar R}_{\ell,k}
&=\left(1-\frac{\tau_{\rm u}+ \tau_{\rm d}}{\tau_{\rm c}}\right)\mathbb{E}_{{{\hat{\mathcal{H}}}_{\ell}}}\left[ R_{\ell,k}\left({\hat{\mathcal{H}}}_{\ell}\right)  \right]\nonumber\\
&=\left(1-\frac{\tau_{\rm u}+ \tau_{\rm d}}{\tau_{\rm c}}\right)\mathbb{E}\left[\log_2\left(1+{\rm SINR}_{\ell,k}\right)\right], \label{eq:ergodic_SE}
\end{align}
where the pre-log term is a normalization factor by the uplink and downlink channel training overhead. To maximize the ergodic spectral efficiency,  we need to optimize the precoding vectors that maximize the {instantaneous} spectral efficiency using noisy CSIT knowledge in every fading state.

\section{Noncooperative Multi-Cell Precoding}
 In this section, we present a novel noncooperative precoding method using local CSIT. To highlight the idea, we focus on a two-cell MU-MIMO system when each BS has perfect knowledge of local CSIT.  

 \subsection{From Centralized to Distributed Precoding}


We commence by reviewing the multi-cell cooperative precoding method using global CSIT \cite{choi2019joint}.  We then explain when the proposed SILNR maximization precoding using local CSIT can achieve the identical performance to the multi-cell cooperative one.



 {\bf Cooperative precoding using global CSIT:} Let ${\bf f}_{\ell}=\left[{\bf f}_{\ell,1}^{\sf H},{\bf f}_{\ell, 2}^{\sf H},\ldots, {\bf f}_{\ell, K}^{\sf H}\right]^{\sf H} \in \mathbb{C}^{N_{\ell}K_{\ell}\times 1}$ be the concatenated precoding vector used at the $\ell$th BS where $\ell\in\{1,2\}$.  We also let ${\bf e}_k=[0,\ldots, 1,\ldots,0]^{\sf T}\in \mathbb{R}^{K_{\ell}\times 1}$ be a unit vector with the nonzero value in the $k$th element. Using this stacked precoding vector, we rewrite the  SINR of the $k$th user in cell $\ell\in \{1,2\}$ in \eqref{eq:SINR} as
\begin{align}
	  {\sf SINR}_{\ell,k}({\bf f}_1,{\bf f}_2) = \frac{{\bf f}_{\ell}^{\sf H}{\bf S}_{\ell,\ell,k}{\bf f}_{\ell} }{{\bf f}_{\ell}^{\sf H}{\bf U}_{\ell,\ell,k}{\bf f}_{\ell}+{\bf f}_{\bar \ell}^{\sf H}{\bf C}_{\bar \ell,\ell,k}{\bf f}_{\bar \ell}},
\end{align}
where $\bar \ell = 3-\ell\in\{1,2\}$, ${\bf S}_{\ell,\ell,k}$, ${\bf U}_{\ell,\ell,k}$, and ${\bf C}_{\bar \ell,\ell,k}$ are defined as
\begin{align}
    {\bf S}_{\ell,\ell,k} &={\bf e}_{k}{\bf e}_{k}^{\sf T} \otimes {\bf h}_{\ell,\ell,k}{\bf h}_{\ell,\ell,k}^{\sf H}\in \mathbb{C}^{N_{\ell}K_{\ell} \times N_{\ell}K_{\ell}},\\
    {\bf U}_{\ell,\ell,k} &= {\bf I}_{K_{\ell}} \otimes {\bf h}_{\ell,\ell,k}{\bf h}_{\ell,\ell,k}^{\sf H} -  {\bf S}_{\ell,\ell,k}+\frac{\sigma^2}{P_{\ell}}{\bf I}_{N_{\ell}K_{\ell}}\in \mathbb{C}^{N_{\ell}K_{\ell} \times N_{\ell}K_{\ell} },\\
    {\bf C}_{\bar \ell,\ell,k} &= {\bf I}_{K_{\ell}} \otimes \frac{P_{\ell}}{P_{\bar\ell}}{\bf h}_{\bar \ell,\ell,k}{\bf h}_{\bar \ell,\ell,k}^{\sf H}\in \mathbb{C}^{N_{\ell}K_{\ell} \times N_{\ell}K_{\ell}}.
\end{align} 
 Then, the sum-spectral efficiency is
\begin{align}
	 R^{\sf sum}({\bf f}_1,{\bf f}_2)= R_1^{\sf sum}({\bf f}_1,{\bf f}_2) + R_2^{\sf sum}({\bf f}_1,{\bf f}_2),
	 \label{eq:coopbound}
\end{align}
where 
\begin{align}
	  R_{\ell}^{\sf sum}({\bf f}_1,{\bf f}_2) &=  \sum_{k=1}^{K_{\ell}}  \log_2\left(   \frac{{\bf f}_{\ell}^{\sf H}\left({\bf S}_{\ell,\ell,k}+{\bf U}_{\ell,\ell,k} \right){\bf f}_{\ell}+{\bf f}_{\bar \ell}^{\sf H} {\bf C}_{\bar \ell,\ell,k} {\bf f}_{\bar \ell} }{ {\bf f}_{\ell}^{\sf H}{\bf U}_{\ell,\ell,k}{\bf f}_{\ell} + {\bf f}_{\bar \ell}^{\sf H} {\bf C}_{\bar \ell,\ell,k} {\bf f}_{\bar \ell}} \right)
	  \end{align}
	  for $\ell\in \{1,2\}$.
Consequently, the sum-spectral efficiency maximization problem is a form: 
\begin{subequations}
\begin{align}
    &{\underset{{\bf f}_{\ell} \in \mathbb{C}^{N_{\ell}K_{\ell}}}{\text{arg~max}}} 
    R^{\sf sum}({\bf f}_1,{\bf f}_2)  
 ,\\
    &\text{subject to} ~~ \|{\bf f}_{\ell}\|_2^2= 1,~~~\forall\ell \in \{1,2\}.  
\end{align}\label{eq:Objective22}
\end{subequations}
This optimization problem finds a joint solution for a set of scheduled users per cell, the precoding vector, and the power allocation per stream for both BSs to maximize the sum-spectral efficiency. Unfortunately, finding a global optimal solution is infeasible in practice. Besides, global CSIT knowledge is required to obtain a local-optimal solution, as shown in \cite{choi2019joint}. This global CSIT knowledge requirement underrates the cooperative transmission gains because of the CSIT sharing overheads.  This is especially pronounced when the number of cooperative BS increases \cite{lee2014spectral,park2015cooperative,park2016optimal}. 

 {\bf Distributed precoding using local CSIT:}
We propose a novel distributed precoding strategy that harnesses local CSIT only.   The central idea is to maximize the sum-spectral efficiency under the local CSIT constraint. To accomplish this, we introduce a new metric named SILNR. The SILNR of the $k$th user of BS $\ell$ is defined as
\begin{align}
	 {\sf SILNR}_{{\ell},k}({\bf f}_{\ell}) = \frac{{\bf f}_{\ell}^{\sf H}{\bf S}_{\ell,\ell,k}{\bf f}_{\ell} }{{\bf f}_{\ell}^{\sf H}{\bf U}_{\ell,\ell,k}{\bf f}_{\ell}+ {\bf L}_{\ell,\bar \ell}({\bf f}_{\ell})^{\frac{K_{\bar \ell}}{K_{\ell}}}  }, \label{eq:SILNR1}
\end{align} 
where ${\bf L}_{\ell,\bar \ell}({\bf f}_{\ell})$ is the geometric mean of the interference leakage to users in the other cells by the transmission of BS $\ell$, i.e.,
\begin{align}
	{\bf L}_{\ell,\bar \ell}({\bf f}_{\ell})=\left(\prod_{j=1}^{K_{\bar \ell}}{\bf f}_{\ell}^{\sf H}{\bf C}_{\ell,\bar \ell,j}{\bf f}_{\ell}\right)^{\frac{1}{K_{\bar \ell}}}. 
\end{align}

%



We provide some remarks on this SILNR value. 
 
 \begin{itemize}
 	\item ${\sf SILNR}_{\ell,k}({\bf f}_{\ell}) $ is a function of only ${\bf f}_{\ell}$ for $\ell\in \{1,2\}$;  this implies that each BS constructs the aggregated multi-user precoding vector ${\bf f}_{\ell}\in \mathbb{C}^{N_{\ell}K_{\ell}}$ without sharing CSIT. 
 	\item When $K_{\bar \ell}=K_{\ell}$, the SILNR takes into account the effective interference leakage power as the geometric mean of $\left\{{\bf f}_{\ell}^{\sf H}{\bf C}_{\ell,\bar \ell,1}{\bf f}_{\ell},\ldots, {\bf f}_{\ell}^{\sf H}{\bf C}_{\ell,\bar \ell,K_{\bar\ell}}{\bf f}_{\ell}\right\}$. This geometric mean structure plays a key role in maximizing the sum-spectral efficiency using local CSIT, which will be explained in the sequel. In addition, when $K_{\bar \ell}\ne K_{\ell}$, the exponent of ${\bf L}_{\ell,\bar\ell}\left({\bf f}_{\ell}\right)$, $\frac{K_{\bar\ell}}{K_{\ell}}$, controls the IUI and the interference leakage toward other cells. For example, in the case when $K_{\ell}\gg K_{\bar\ell}$, this exponent $\frac{K_{\bar\ell}}{K_{\ell}}$ diminishes the interference leakage power. This implies that the precoder devotes to reduce the IUI power more than the leakage interference power.
 	\item ${\sf SILNR}_{\ell,k}({\bf f}_{\ell})$ is always smaller than $ \rho_{\ell,k}({\bf f}_{\ell}) = \frac{{\bf f}_{\ell}^{\sf H}{\bf S}_{\ell,\ell,k}{\bf f}_{\ell} }{{\bf f}_{\ell}^{\sf H}{\bf U}_{\ell,\ell,k}{\bf f}_{\ell}}$, because the leakage power term $ {\bf L}_{\ell,\bar \ell}({\bf f}_{\ell})^{\frac{K_{\bar \ell}}{K_{\ell}}} $ is positive.  Therefore, one can interpret a function $	\log_2\left(1+{\sf SILNR}_{\ell,k}\right)$ as a lower bound of the spectral efficiency of $\log_2\left(1+\rho_{\ell,k}\right)$ by penalizing the interference leakage generated by BS $\ell$.  	

  
 \end{itemize}

Using this new metric, we define a sum-rate function of BS $\ell$ as

\begin{align}
&{\hat R}_{\ell}^{\sf sum}({\bf f}_{\ell}) = \sum_{k=1}^{K_{\ell}}\log_2\left( 1+  {\sf SILNR}_{\ell,k}({\bf f}_{\ell}) \right) \nonumber\\
&= \log_2\left(   \prod_{k=1}^{K_{\ell}} \frac{ {\bf f}_{\ell}^{\sf H}\left({\bf S}_{\ell,\ell,k}+{\bf U}_{\ell,\ell,k} \right){\bf f}_{\ell}+{\bf L}_{\ell,\bar \ell}({\bf f}_{\ell})^{\frac{K_{\bar \ell}}{K_{\ell}}} }{  {\bf f}_{\ell}^{\sf H}{\bf U}_{\ell,\ell,k}{\bf f}_{\ell}+ {\bf L}_{\ell,\bar \ell}({\bf f}_{\ell})^{\frac{K_{\bar \ell}}{K_{\ell}}} } \right).\label{eq:lower_bound}
\end{align}

Consequently, each BS independently identifies a joint solution for the user-selection, precoding, and power allocation by exploiting local CSIT. The optimization problem is the following form:
\begin{subequations}
\begin{align}
    &{\underset{{\bf f}_{\ell} \in \mathbb{C}^{N_{\ell}K_{\ell}\times 1} }{\text{arg~max}}} 
  \sum_{k=1}^{K_{\ell}} \log_2\left(1+ {\sf SILNR}_{\ell,k}({\bf f}_{\ell})\right),\\ 
      &\text{subject to} ~~\| {\bf f}_{\ell}\|_2^2=1,~~~\forall k \in{\mathcal{K}_{\ell}}.
\end{align}\label{eq:SILNRmax}
\end{subequations}


\subsection{Quasi-Optimal Cases}
To shed light on the idea, it is instructive to consider two cases in which the proposed distributed precoding method achieves the cooperative precoding bound very closely.   

{\bf Case 1 (Zero-IUI condition):}  Let us consider a multi-cell cooperative precoding strategy that maximizes the tight lower bound under zero-IUI constraint.  From \eqref{eq:Objective22}, the sum-spectral efficiency maximization problem under the zero-IUI constraint becomes
\begin{subequations}
\begin{align}
    &{\underset{{\bf f}_{\ell} \in \mathbb{C}^{N_{\ell}K_{\ell} }}{\text{arg~max}}} 
  \sum_{\ell=1}^2\sum_{k=1}^{K_{\ell}} \log_2\left(1+  {\sf SINR}_{\ell,k}({\bf f}_1,{\bf f}_2)\right)
    ,\label{eq:Objective232}\\
    &\text{subject to} ~~{\bf f}_{\ell}^{\sf H}{\bf U}_{\ell,k}{\bf f}_{\ell}=0,~~~\forall\ell \in \{1,2\},\forall k\in{\mathcal{K}_{\ell}},\\
            &~~~~~~~~~~~~~ \|{\bf f}_{\ell}\|_2^2= 1,~~~\forall\ell \in \{1,2\}.  
\end{align} \label{eq:zeroIUI}
\end{subequations}
This optimization finds a joint solution for a set of scheduled users per cell, the precoding vector, and the power allocation per stream for both BSs.  Since $ \log_2\left( 1+ {\sf SINR}_{\ell,k}({\bf f}_1,{\bf f}_2)\right) \simeq \log_2\left({\sf SINR}_{\ell,k}({\bf f}_1,{\bf f}_2)\right) $ for a high SINR regime, when ${\bf f}_{\ell}^{\sf H}{\bf U}_{\ell,k}{\bf f}_{\ell}=0$, we approximate the sum-spectral efficiency as  
\begin{align}
	&R^{\sf sum}({\bf f}_1,{\bf f}_2)\simeq \sum_{\ell=1}^2\sum_{k=1}^{K_{\ell}} \log_2\left(  {\sf SINR}_{\ell,k}({\bf f}_1,{\bf f}_2)\right)\nonumber\\
	&= \log_2\left(  \prod_{k=1}^{K_1}  \frac{{\bf f}_{1}^{\sf H}{\bf S}_{1,1,k}{\bf f}_{1} }{ {\bf f}_{2}^{\sf H}{\bf C}_{2,1,k}{\bf f}_{2}} \prod_{k=1}^{K_2} \frac{{\bf f}_{2}^{\sf H}{\bf S}_{2,2,k}{\bf f}_{2} }{ {\bf f}_{1}^{\sf H}{\bf C}_{1,2,k}{\bf f}_{1}  } \right)\nonumber\\
	&=   \log_2\left(\frac{\prod_{k=1}^{K_1}{\bf f}_{1}^{\sf H}{\bf S}_{1,1,k}{\bf f}_{1} }{\prod_{k=1}^{K_2} {\bf f}_{1}^{\sf H}{\bf C}_{1,2,k} {\bf f}_{1}}  \right) +  \log_2\left( \frac{\prod_{k=1}^{K_2}{\bf f}_{2}^{\sf H}{\bf S}_{2,2,k}{\bf f}_{2} }{\prod_{k=1}^{K_1}{\bf f}_{2}^{\sf H}{\bf C}_{2,1,k}{\bf f}_{2}  } \right).
\end{align}
Thanks to the zero-IUI condition, the sum-spectral efficiency maximization problem becomes separable with respect to each optimization vector. Therefore, the solution for the joint optimization problem for the aggregated precoding vectors ${\bf f}_{\ell}$ is obtained by solving two separate optimization problems independently:
\begin{subequations}
\begin{align}
    &{\underset{{\bf f}_{\ell} \in \mathbb{C}^{N_{\ell}K_{\ell}\times 1}}{\text{arg~max}}} 
   \log_2\left(\frac{\prod_{k=1}^{K_{\ell}}{\bf f}_{\ell}^{\sf H}{\bf S}_{\ell,\ell,k}{\bf f}_{\ell} }{\prod_{k=1}^{K_{\bar \ell}}{\bf f}_{\ell}^{\sf H}{\bf C}_{\ell,\bar \ell,k}{\bf f}_{\ell}}  \right)     ,\label{eq:sub1}\\
    &\text{subject to} ~~{\bf f}_{\ell}^{\sf H}{\bf U}_{\ell,\ell,k}{\bf f}_{\ell}=0,~~~\forall k\in {\mathcal{K}_{\ell}}, \\
           &~~~~~~~~~~~~~ \|{\bf f}_{\ell}\|_2^2= 1.   
\end{align}
\end{subequations}

Now we turn our attention to the sum-rate function defined in \eqref{eq:lower_bound}. When ${\bf f}_{\ell}^{\sf H}{\bf U}_{\ell,\ell,k}{\bf f}_{\ell}=0$, we approximate the sum-rate function as
\begin{align}
	&{\hat R}_{\ell}^{\sf sum}({\bf f}_{\ell}) \simeq \sum_{k=1}^{K_{\ell}}\log_2\left(  {\sf SILNR}_{\ell,k}({\bf f}_{\ell}) \right)\nonumber\\ 
	&= \log_2\left(    \frac{\prod_{k=1}^{K_{\ell}} {\bf f}_{\ell}^{\sf H} {\bf S}_{\ell,\ell,k}   {\bf f}_{\ell}  }{\prod_{k=1}^{K_{\ell}}   {\bf L}_{\ell,\bar \ell}({\bf f}_{\ell})^{\frac{K_{\bar \ell}}{K_{\ell}}} } \right)
	= \log_2\left(    \frac{\prod_{k=1}^{K_{\ell}} {\bf f}_{\ell}^{\sf H} {\bf S}_{\ell,\ell,k}  {\bf f}_{\ell}  }{ \prod_{j=1}^{K_{\bar \ell}}{\bf f}_{\ell}^{\sf H}{\bf C}_{\ell,\bar \ell,j}{\bf f}_{\ell} } \right).\label{eq:lower_bound2}
\end{align}
It is remarkable that \eqref{eq:sub1} and \eqref{eq:lower_bound2} are identical. We conclude that the proposed distributed precoding using local CSIT asymptotically achieves the cooperative bound in the high SINR regime under the zero-IUI condition.  This equivalence comes from our SILNR definition using the geometric mean of the interference leakage terms. 


{\bf Case 2 (Zero-ICI condition):} Let us consider the zero-ICI constraint in both sum-spectral efficiency maximization problems in \eqref{eq:Objective22} and \eqref{eq:lower_bound}. When $N_{\ell}>K_{\bar\ell}$, it is possible to meet the zero-ICI condition by constructing precoding vectors on the nullspace of the column space spanned by the ICI channels.  In this manner, we begin by reformulating the problem in \eqref{eq:Objective22} under zero-ICI constraint as
\begin{subequations}
\begin{align}
    &{\underset{{\bf f}_{\ell} \in \mathbb{C}^{N_{\ell}K_{\ell} }}{\text{arg~max}}} 
  \sum_{\ell=1}^2\sum_{k=1}^{K_{\ell}} \log_2\left(1+  {\sf SINR}_{\ell,k}({\bf f}_1,{\bf f}_2)\right)
    ,\label{eq:Objective2}\\
    &\text{subject to} ~~{\bf f}_{\ell}^{\sf H}{\bf C}_{\ell,\bar\ell,k}{\bf f}_{\ell}=0,~~~\forall\ell \in \{1,2\},\forall k\in{\mathcal{K}_{\ell}},\\
            &~~~~~~~~~~~~~ \|{\bf f}_{\ell}\|_2^2= 1,~~~\forall\ell \in \{1,2\}.  
\end{align}
\end{subequations}

Under the zero-ICI constraint, the sum-spectral efficiency simplifies  
\begin{align}
	&\sum_{\ell=1}^2\sum_{k=1}^{K_{\ell}} \log_2\left(1+  {\sf SINR}_{\ell,k}({\bf f}_1,{\bf f}_2)\right)\nonumber\\
	&=\log_2\left(\prod_{k=1}^{K_1} \frac{{\bf f}_{1}^{\sf H}\left( {\bf S}_{1,1,k}+{\bf U}_{1,1,k}\right){\bf f}_{1} }{ {\bf f}_{1}^{\sf H}{\bf U}_{1,1,k} {\bf f}_{1}}  \right)\nonumber\\
	&+\log_2\left(\prod_{k=1}^{K_2}   \frac{{\bf f}_{2}^{\sf H}\left({\bf S}_{2,2,k}+{\bf U}_{2,2,k}\right){\bf f}_{2} }{ {\bf f}_{2}^{\sf H} {\bf U}_{2,2,k} {\bf f}_{2}  }\right).
\end{align}
This zero-ICI condition also makes the joint precoding design problem separable as
\begin{subequations}
\begin{align}
    &{\underset{{\bf f}_{\ell} \in \mathbb{C}^{N_{\ell}K_{\ell}}}{\text{arg~max}}} 
   \log_2\left(   \prod_{k=1}^{K_{\ell}} \frac{{\bf f}_{\ell}^{\sf H}\left({\bf S}_{\ell,\ell,k}+{\bf U}_{\ell,\ell,k}\right){\bf f}_{\ell} }{ {\bf f}_{\ell}^{\sf H} {\bf U}_{\ell,\ell,k}  {\bf f}_{\ell}}  \right)     ,\label{eq:sub3}\\
    &\text{subject to} ~~{\bf f}_{\ell}^{\sf H}{\bf C}_{\ell,\bar \ell,k}{\bf f}_{\ell}=0,~~~\forall k\in {\mathcal{K}_{\ell}},\\  
           &~~~~~~~~~~~~~ \|{\bf f}_{\ell}\|_2^2= 1.
\end{align}
\end{subequations}

From  \eqref{eq:lower_bound}, when ${\bf L}_{\ell,\bar \ell}({\bf f}_{\ell})=0$, our objective function becomes
\begin{align}
{\hat R}_{\ell}^{\sf sum}({\bf f}_{\ell}) = \log_2\left(  \prod_{k=1}^{K_{\ell}} \frac{ {\bf f}_{\ell}^{\sf H}\left({\bf S}_{\ell,\ell,k}+{\bf U}_{\ell,\ell,k} \right){\bf f}_{\ell} }{    {\bf f}_{\ell}^{\sf H}{\bf U}_{\ell,\ell,k}{\bf f}_{\ell}  } \right).\label{eq:app2}
\end{align}
This result implies that our precoding strategy exploiting local CSIT is sufficient to achieve the multi-cell cooperation bound very closely under either zero-IUI or zero-ICI constraints.  This result, however, holds for the two-cell scenario only. In the sequel, we generalize this idea in a general multi-cell scenario and noisy CSIT assumption, which are more practically relevant.

\section{Multi-Cell SILNR Maximization Precoding with Noisy CSIT}
This section generalizes the idea of SILNR maximization precoding introduced in the previous section to a multi-cell noisy CSIT scenario. We start with formulating the SILNR maximization problem under noisy CSIT setting. Then, we establish the first- and second-order optimality condition for the problem. Then, a computationally-efficient algorithm to find a local-optimal solution to the problem is presented.

\subsection{Problem Formulation}
Using noisy and local CSIT at the $\ell$th BS, the received signal of the $k$th user in the $\ell$th cell is rewritten as
\begin{align}
    y_{\ell,k}[t]&={\bf \hat h}_{\ell,\ell,k}^{\sf H}{\bf f}_{\ell,k}{s_{\ell,k}}[t]
    +\sum_{i\ne k}^{K_{\ell}}{\bf \hat h}_{\ell,\ell,k}^{\sf H}{\bf f}_{\ell,i}{s_{\ell,i}}[t]\nonumber\\
    &+\sum_{i=1}^{K_{\ell}}{\bf e}_{\ell,\ell,k}^{\sf H}{\bf f}_{\ell, i}s_{\ell,i}[t]
    +{\tilde n}_{\ell,k}[t],\label{eq:effective_recv}
\end{align}
where  ${\tilde n}_{\ell,k}[t]=\sum_{j\ne\ell}\sum_{k=1}^{K_j}{\bf h}_{j,\ell,k}^{\sf H}{\bf f}_{j,k}s_{j,k}[t]\!+\!n_{\ell,k}[t]$ is the effective noise when treating all aggregated ICI as additional noise and distributed by ${\tilde n}_{\ell,k}[t]\sim\mathcal{CN}\left(0,\tilde\sigma_{\ell,k}^2\right)$, where $\tilde\sigma_{\ell,k}^2=\mathbb{E}[\sum_{j\ne\ell}\sum_{k=1}^{K_j}|{\bf h}_{j,\ell,k}^{\sf H}{\bf f}_{j,k}s_{j,k}[t]|^2]+\sigma^2$. We define the geometric mean of the interference leakage in \eqref{eq:SILNR1} for a multi-cell scenario. Let $\mathcal{U}_{\ell}$ be the collection of the other cell's users who use the orthogonal pilots with the users in the $\ell$th cell, i.e., $\mathcal{U}_{\ell}=\left\{(\bar\ell,j)~|~\forall\bar\ell\in\mathcal{C}_{\ell}\setminus \left\{\ell\right\}, \forall j\in\mathcal{K}_{\bar\ell}\right\}$. Then, the leakage interference ${\bf\hat L}_{\ell}({\bf f}_{\ell})$ becomes 
\begin{align}
	{\bf\hat L}_{\ell}({\bf f}_{\ell})
	=\left(\prod_{(\bar\ell,j)\in\mathcal{U}_{\ell}}{\bf f}_{\ell}^{\sf H}{\bf\hat C}_{\ell,\bar \ell,j}{\bf f}_{\ell}\right)^{\frac{1}{\left|\mathcal{U}_{\ell}\right|}},  \label{eq:leakage}
\end{align}
where ${\bf\hat C}_{\ell,\bar \ell,j}={\bf I}_{K_{\ell}} \otimes \frac{P_{\bar \ell}}{P_{\ell}}\left({\bf \hat h}_{\ell,\bar \ell,j}{\bf\hat h}_{\ell,\bar \ell,j}^{\sf H}+{\bf \Phi}_{\ell,\bar \ell,j}\right)\in \mathbb{C}^{N_{\ell}K_{\ell} \times N_{\ell}K_{\ell}}$.
Incorporating \eqref{eq:effective_recv} and \eqref{eq:leakage}, we define SILNR of the $k$th user in the $\ell$th cell as 
\begin{align}
    &{\sf SILNR}_{{\ell},k}({\bf f}_{\ell})\nonumber\\
    &=\frac{\left|{\bf \hat h}_{\ell,\ell,k}^{\sf H}{\bf f}_{\ell,k}\right|^2}{\sum_{i\ne k}^{K_{\ell}}\left|{\bf\hat h}_{\ell,\ell,k}^{\sf H}{\bf f}_{\ell,i}\right|^2+\sum_{i=1}^{K_{\ell}}{\bf f}_{\ell,i}^{\sf H}{\bf \Phi}_{\ell,\ell,k}{\bf f}_{\ell,i}+{\bf\hat L}_{\ell}({\bf f}_{\ell})^{\frac{\left|\mathcal{U}_{\ell}\right|}{K_{\ell}}}+\frac{{\tilde\sigma}_{\ell,k}^2}{P_{\ell}}}\nonumber\\
    &= \frac{{\bf f}_{\ell}^{\sf H}{\bf\hat S}_{\ell,\ell,k}{\bf f}_{\ell} }
    {{\bf f}_{\ell}^{\sf H}{\bf\hat U}_{\ell,\ell,k}{\bf f}_{\ell} 
    + {\bf f}_{\ell}^{\sf H}{\bf E}_{\ell,\ell,k}{\bf f}_{\ell} 
    + {\bf\hat L}_{\ell}({\bf f}_{\ell})^{\frac{\left|\mathcal{U}_{\ell}\right|}{K_{\ell}}}  }, \label{eq:SILNR_Multi_Cell}
\end{align}
where
\begin{align}
    {\bf\hat S}_{\ell,\ell,k} &={\bf e}_{k}{\bf e}_{k}^{\sf T} \otimes {\bf \hat h}_{\ell,\ell,k}{\bf\hat h}_{\ell,\ell,k}^{\sf H}\in \mathbb{C}^{N_{\ell}K_{\ell} \times N_{\ell}K_{\ell}}, \nonumber\\
    {\bf\hat U}_{\ell,\ell,k} &= {\bf I}_{K_{\ell}} \otimes {\bf\hat h}_{\ell,\ell,k}{\bf\hat h}_{\ell,\ell,k}^{\sf H}\!-\!{\bf\hat S}_{\ell,\ell,k}\!+\!\frac{\tilde\sigma_{\ell,k}^2}{P_{\ell}}{\bf I}_{N_{\ell}K_{\ell}}\in \mathbb{C}^{N_{\ell}K_{\ell} \times N_{\ell}K_{\ell} }, \nonumber\\
    {\bf E}_{\ell,\ell,k} &= {\bf I}_{K_{\ell}} \otimes {\bf\Phi}_{\ell,\ell,k}\in \mathbb{C}^{N_{\ell}K_{\ell} \times N_{\ell}K_{\ell}}.
\end{align} 
Accordingly, our precoding strategy using noisy and local CSIT is to solve the following optimization problem:
\begin{subequations}
\label{eq:Problem_Rayleigh_Quotients}
\begin{align}
    &{\underset{{\bf f}_{\ell}\in \mathbb{C}^{N_{\ell}K_{\ell} }}{\text{arg~max}}} \prod_{k=1}^{K_{\ell}}
    \frac{{\bf f}_{\ell}^{\sf H}\left({\bf\hat S}_{\ell,\ell,k}\!+\!{\bf\hat U}_{\ell,\ell,k}\!+\!{\bf E}_{\ell,\ell,k}
    \!+\!{\bf\hat L}_{\ell}({\bf f}_{\ell})^{\frac{\left|\mathcal{U}_{\ell}\right|}{K_{\ell}}}\right){\bf f}_{\ell}}
    {{\bf f}_{\ell}^{\sf H}\left({\bf\hat U}_{\ell,\ell,k}\!+\!{\bf E}_{\ell,\ell,k}\!+\!{\bf\hat L}_{\ell}({\bf f}_{\ell})^{\frac{\left|\mathcal{U}_{\ell}\right|}{K_{\ell}}}\right){\bf f}_{\ell}}
    ,\label{eq:Product_Rayleigh} \\
    &\text{subject to} ~~ \|{\bf f}_{\ell}\|_2^2= 1,~~~\forall\ell \in \mathcal{L}. \label{eq:Relaxed_Constraint}
\end{align}
\end{subequations}
Since the objective function \eqref{eq:Problem_Rayleigh_Quotients} is highly non-convex, finding even local-optimal solution for ${\bf f}_{\ell}$ is a very challenging task. In the sequel, we derive the first- and second-order optimality conditions for this non-convex optimization problem. 

\subsection{Local Optimality Conditions}

The following theorems establish the first- and the second-order necessary conditions for the local optimality of the non-convex optimization problem in \eqref{eq:Problem_Rayleigh_Quotients}.

\begin{thm}
\label{thm1}
{\bf(The first-order necessary condition)} If ${\bf f}_{\ell}^{\star}\in \mathbb{C}^{N_{\ell}K_{\ell}\times 1}$ is a stationary point of the non-convex optimization problem \eqref{eq:Problem_Rayleigh_Quotients}, it satisfies  
\begin{align}
    {\bf \bar A}_{\ell}\left({\bf f}_{\ell}^{\star}\right){\bf f}_{\ell}^{\star}=
    \gamma\left({\bf f}_{\ell}^{\star}\right){\bf \bar B}_{\ell}\left({\bf f}_{\ell}^{\star},\lambda\right){\bf f}_{\ell}^{\star},\label{eq:cond1}
\end{align}
where the functional matrices ${\bf \bar A}_{\ell}\left({\bf f}_{\ell}^{\star}\right)$ and ${\bf \bar B}_{\ell}\left({\bf f}_{\ell}^{\star},\lambda\right)$ are 
\begin{align}
    &{\bf\tilde A}_{\ell,\ell,k} = {\bf\hat S}_{\ell,\ell,k}\!+\!{\bf\hat U}_{\ell,\ell,k}\!+\!{\bf E}_{\ell,\ell,k},\nonumber\\
    &{\bf\tilde B}_{\ell,\ell,k} = {\bf\hat U}_{\ell,\ell,k}\!+\!{\bf E}_{\ell,\ell,k},\nonumber
\end{align}
\begin{align}
    &{\bf \bar A}_{\ell}\left({\bf f}_{\ell}\right)
    =\sum_{i=1}^{K_{\ell}}\left(\prod_{k\ne i }^{K_{\ell}}({\bf f}_{\ell})^{\sf H}\left({\bf \tilde A}_{\ell,\ell,k}+{\bf\hat L}_{\ell}({\bf f}_{\ell})^{\frac{\left|\mathcal{U}_{\ell}\right|}{K_{\ell}}}\right){\bf f}_{\ell}\right)\nonumber\\
    &~~~~~~~~\times\left({\bf \tilde A}_{\ell,\ell,i}+{\bf\tilde C}_{\ell,\ell,k}\right){\bf f}_{\ell},\nonumber\\
    &{\bf \bar B}_{\ell}\left({\bf f}_{\ell},\lambda\right)
    =\left[\sum_{i=1}^{K_{\ell}}\left(\prod_{k\ne i}^{K_{\ell}}{({\bf f}_{\ell})^{\sf H}\left({\bf \tilde B}_{\ell,\ell,k}\!+\!{\bf\hat L}_{\ell}({\bf f}_{\ell})^{\frac{\left|\mathcal{U}_{\ell}\right|}{K_{\ell}}}\right){\bf f}_{\ell}}\right)\right.\nonumber\\
    &~~~~~~~~~~~\left.\times\left({\bf \tilde{B}_{\ell,\ell,i}}
    +{\bf\tilde C}_{\ell,\ell,k}\right)+\frac{\lambda}{\gamma({\bf f}_{\ell})}{\bf I}_{N_{\ell}K_{\ell}}\right]{\bf f}_{\ell},\nonumber\\
    &{\bf\tilde C}_{\ell,\ell,k} = \sum_{(\bar\ell,j)\in\mathcal{U}_{\ell}}\frac{{\bf\hat L}_{\ell}({\bf f}_{\ell})^{\frac{\left|\mathcal{U}_{\ell}\right|}{K_{\ell}}}{\bf\hat C}_{\ell,\bar \ell,j}}{K_{\ell}({\bf f}_{\ell})^{\sf H}{\bf\hat C}_{\ell,\bar \ell,j}{\bf f}_{\ell}},\nonumber\\
    &\gamma({\bf f}_{\ell})=\prod_{k=1}^{K_{\ell}}
    \frac{{\bf f}_{\ell}^{\sf H}\left({\bf\hat S}_{\ell,\ell,k}\!+\!{\bf\hat U}_{\ell,\ell,k}\!+\!{\bf E}_{\ell,\ell,k}
    \!+\!{\bf\hat L}_{\ell}({\bf f}_{\ell})^{\frac{\left|\mathcal{U}_{\ell}\right|}{K_{\ell}}}\right){\bf f}_{\ell}}
    {{\bf f}_{\ell}^{\sf H}\left({\bf\hat U}_{\ell,\ell,k}\!+\!{\bf E}_{\ell,\ell,k}\!+\!{\bf\hat L}_{\ell}({\bf f}_{\ell})^{\frac{\left|\mathcal{U}_{\ell}\right|}{K_{\ell}}}\right){\bf f}_{\ell}}.
\label{eq:matrices}
\end{align}
In addition, the Lagrange multiplier $\lambda$ is chosen so that ${\bf f}$ satisfies $\left\|{\bf f}_{\ell}\right\|_2^2=1$.
\end{thm}
\begin{IEEEproof}
See Appendix A.
\end{IEEEproof}
\vspace{0.3cm}

Theorem \ref{thm1} implies that any stationary point of the non-convex problem in \eqref{eq:Problem_Rayleigh_Quotients} is one of the eigenvectors of the functional matrix $[{\bf \bar B}_{\ell}\left({\bf f}_{\ell},\lambda\right)]^{-1}{\bf \bar A}_{\ell}$, i.e., 
\begin{align}
	[{\bf \bar B}_{\ell}\left({\bf f}_{\ell},\lambda\right)]^{-1}{\bf \bar A}_{\ell}\left({\bf f}_{\ell}\right){\bf f}_{\ell}=
    \gamma\left({\bf f}_{\ell}\right){\bf f}_{\ell}.
\end{align}
As can be seen, the objective function $\gamma\left({\bf f}_{\ell}\right)$ can be interpreted as the eigenvalue of the functional matrix $[{\bf \bar B}_{\ell}\left({\bf f}_{\ell},\lambda\right)]^{-1}{\bf \bar A}_{\ell}$. Since we are interested in maximizing the objective function $\gamma\left({\bf f}_{\ell}\right)$, we need to identify the eigenvector corresponding to the maximum eigenvalue, which can be a global optimal solution. Unfortunately, finding such eigenvector is highly non-trivial, because $[{\bf \bar B}_{\ell}\left({\bf f}_{\ell},\lambda\right)]^{-1}{\bf \bar A}_{\ell}$ is a function of ${\bf f}_{\ell}$ and Lagrange multiplier $\lambda$. Nevertheless, this principal component analysis helps to understand the global landscape of the non-convex optimization problem; and leads to an algorithm to find a local-optimal solution in a numerically efficient manner.

Although ${\bf f}^{\star}_{\ell}$ satisfies the first-order necessary condition derived in Theorem \ref{thm1}, we need to check the curvature of the objective function around the stationary point to verify the local optimality.  The following theorem gives a testing condition of the  negative definiteness of the extended Hessian matrix evaluated at ${\bf f}^{\star}_{\ell}$ in a closed-form, i.e., $\nabla_{{\bf f}_{\ell}^{\sf H}}^2\gamma({\bf f}_{\ell}^{\star})\prec 0$. 

\begin{figure*}[t]
\begin{align}
&\!\!\!\!\!\!\!\!\!\lambda_{\sf max}
\left(\left[
    \sum_{i=1}^{K_{\ell}}
    \frac{{\bf X}_{\ell,\ell,i}^{A}{\bf f}_{\ell}}
    {{\bf f}_{\ell}^{\sf H}{\bf Y}_{\ell,\ell,i}^{A}{\bf f}_{\ell}}-
    \frac{{\bf X}_{\ell,\ell,i}^{B}{\bf f}_{\ell}}
    {{\bf f}_{\ell}^{\sf H}{\bf Y}_B{\bf f}_{\ell}}
    \right]
    \left[
    \sum_{i=1}^{K_{\ell}}
    \frac{{\bf X}_{\ell,\ell,i}^{A}{\bf f}_{\ell}}
    {{\bf f}_{\ell}^{\sf H}{\bf Y}_{\ell,\ell,i}^{A}{\bf f}_{\ell}}-
    \frac{{\bf X}_{\ell,\ell,i}^{B}{\bf f}_{\ell}}
    {{\bf f}_{\ell}^{\sf H}{\bf Y}_B{\bf f}_{\ell}}
    \right]^{\sf H}
    +\gamma({\bf f}_{\ell})
    \left[
    \sum_{i=1}^{K_{\ell}}
    \frac{{\bf X}_{\ell,\ell,i}^{A}}{{\bf f}_{\ell}^{\sf H}{\bf Y}_{\ell,\ell,i}^A{\bf f}_{\ell}}
    +\frac{{\bf Z}_{\ell,\ell,i}}{{\bf f}_{\ell}^{\sf H}{\bf Y}_{\ell,\ell,i}^B{\bf f}_{\ell}}
    +\frac{{\bf X}_{\ell,\ell,i}^B{\bf f}_{\ell}{\bf f}_{\ell}^{\sf H}{\bf X}_{\ell,\ell,i}^B}{\left({\bf f}_{\ell}^{\sf H}{\bf Y}_{\ell,\ell,i}^B{\bf f}_{\ell}\right)^2}\right]
\right)\nonumber\\
&<
\lambda_{\sf min}
\left(
\gamma({\bf f}_{\ell})
    \left[\sum_{i=1}^{K_{\ell}}
    \frac{{\bf X}_{\ell,\ell,i}^{B}}{{\bf f}_{\ell}^{\sf H}{\bf Y}_{\ell,\ell,i}^B{\bf f}_{\ell}}
    +\frac{{\bf Z}_{\ell,\ell,i}}{{\bf f}_{\ell}^{\sf H}{\bf Y}_{\ell,\ell,i}^A{\bf f}_{\ell}}
    +\frac{{\bf X}_{\ell,\ell,i}^A{\bf f}_{\ell}{\bf f}_{\ell}^{\sf H}{\bf X}_{\ell,\ell,i}^A}{\left({\bf f}_{\ell}^{\sf H}{\bf Y}_{\ell,\ell,i}^A{\bf f}_{\ell}\right)^2}
    \right]+\lambda{\bf I}
\right).
\label{eq:cond2}    
\end{align}
\hrulefill
\end{figure*}

\begin{thm}
\label{thm2}
{\bf(The second-order necessary condition)}  The stationary point ${\bf f}_{\ell}^{\star}$ is a local-optimal solution, provided that \eqref{eq:cond2}, where the functional matrices are defined as ${\bf Z}_{\ell,\ell,i}=\sum_{(m,n)\in\mathcal{U}_{\ell}}\frac{{\bf\hat L}_{\ell}\left({\bf f}_{\ell}\right)^{\frac{|\mathcal{U}_{\ell}|}{K_{\ell}}}{\bf\hat C}_{\ell,m,n}{\bf f}_{\ell}{\bf f}_{\ell}^{\sf H}{\bf\hat C}_{\ell,m,n}}{K_{\ell}\left({\bf f}_{\ell}^{\sf H}{\bf\hat C}_{\ell,m,n}{\bf f}_{\ell}\right)^2}$, ${\bf X}_{\ell,\ell,i}^{A} = \left({\bf\tilde A}_{\ell,\ell,i}+{\bf\tilde C}_{\ell,\ell,i}\right)$, ${\bf X}_{\ell,\ell,i}^{B} = \left({\bf\tilde B}_{\ell,\ell,i}+{\bf\tilde C}_{\ell,\ell,i}\right)$, ${\bf Y}_{\ell,\ell,i}^{A}={\bf\tilde A}_{\ell,\ell,i}+{\bf\hat L}_{\ell}\left({\bf f}_{\ell}\right)^{\frac{|\mathcal{U}_{\ell}|}{K_{\ell}}}$ and ${\bf Y}_B={\bf\tilde B}_{\ell,\ell,i}+{\bf\hat L}_{\ell}\left({\bf f}_{\ell}\right)^{\frac{|\mathcal{U}_{\ell}|}{K_{\ell}}}$.

\end{thm}
\begin{IEEEproof}
See Appendix B.
\end{IEEEproof}
\vspace{0.3cm}

Theorem \ref{thm2}  is useful when evaluating the second-order optimality condition, because the direct computation of Hessian matrix $\nabla_{{\bf f}_{\ell}^{\sf H}}^2\gamma({\bf f}_{\ell}^{\star})$ is unnecessary. Although Theorem \ref{thm1} and Theorem \ref{thm2} provide a guidance  for the local optimality conditions for problem \eqref{eq:Problem_Rayleigh_Quotients}, finding such a precoding vector is still challenging. To resolve this issue, we propose a computationally-efficient algorithm in the sequel.


\subsection{Algorithm}
\begin{algorithm}[t]
\caption{SILNR Maximization Precoding}
Initialization: $t=n=0,~ {\bf f}_{\ell}^{(0)}={\sf ZF},~ {\bf f}_{\ell}^{(-1)}={\bf 0}, ~\lambda^{(0)},~ {\text{and }} \epsilon$\;
\While{$\|{\bf f}_{\ell}\|_2^2> 1$}
{
    $n\gets n+1$\;
    $\lambda^{(n)}\gets \lambda^{(n-1)}+\Delta\lambda^{(n)}$\;
    \While{$\|{\bf f}_{\ell}^{(t-1)}-{\bf f}_{\ell}^{(t)}\|_2\ge\epsilon$}
    {
        $t\gets t+1$\;
        ${\bf f}_{\ell}^{(t)} \gets \frac{{\bf f}_{\ell}^{(t)}}{\|{\bf f}_{\ell}^{(t)}\|_2}$\;
        ${\bf f}_{\ell}^{(t)} \gets \left[{\bf \bar B}_{\ell}\left({\bf f}_{\ell}^{(t-1)},\lambda^{(n)}\right)\right]^{-1}{\bf \bar A}_{\ell}\left({\bf f}_{\ell}^{(t-1)}\right){\bf f}_{\ell}^{(t-1)}$\;
        
    }
}   
\label{alg:Max_SILNR}
\end{algorithm}

The proposed algorithm takes two steps: 1) the identification of a stationary point using the generalized power iteration (GPI) technique in \cite{choi2019joint} for a given Lagrange multiplier $\lambda^{(n)}$ and 2) the Lagrangian multiplier adjustment to be a feasible solution for a given ${\bf f}_{\ell}^{(t)}$. The algorithm starts with an initial precoding solution ${\bf f}_{\ell}^{(0)}$, which is typically chosen as the ZF beamforming method. In the $t$th iteration, the algorithm evaluates the functional matrices ${\bf \bar A}_{\ell}\left({\bf f}_{\ell}^{(t-1)}\right)$ and $ {\bf \bar B}_{\ell}\left({\bf f}_{\ell}^{(t-1)},\lambda^{(n)}\right)$ defined in \eqref{eq:matrices} using the identified precoding vector in the previous iteration, i.e., ${\bf f}_{\ell}^{(t-1)}$. Then, it is multiplied with $\left[{\bf \bar B}_{\ell}\left({\bf f}_{\ell}^{(t-1)},\lambda^{(n)}\right)\right]^{-1}{\bf \bar A}_{\ell}\left({\bf f}_{\ell}^{(t-1)}\right)$ to obtain ${\bf f}_{\ell}^{(t)}=\left[{\bf \bar B}_{\ell}\left({\bf f}_{\ell}^{(t-1)},\lambda^{(n)}\right)\right]^{-1}{\bf \bar A}_{\ell}\left({\bf f}_{\ell}^{(t-1)}\right){\bf f}_{\ell}^{(t-1)}$. The iteration continues until a stopping condition $\|{\bf f}_{\ell}^{(t-1)}-{\bf f}_{\ell}^{(t)}\|_2\leq \epsilon$ holds, where $\epsilon$ is selected as a small positive number.
Once the eigenvector ${\bf f}_{\ell}$ that satisfies Theorem \ref{thm1} is identified, i.e., the inner loop of Algorithm \ref{alg:Max_SILNR} is converged, Algorithm \ref{alg:Max_SILNR} checks whether the obtained solution is a feasible point or not. If the obtained solution exceeds the power constraint, Algorithm \ref{alg:Max_SILNR} updates the Lagrange multiplier to meet the constraint. The algorithm proceeds until it finds the feasible solution that satisfies the first-order optimality condition.

\vspace{0.2cm} 
{\bf Remark 1 (The computational complexity of the algorithm):} The computational complexity order of the proposed  SILNR maximization precoding algorithm is $\mathcal{O}\left(JN_{\ell}^2K_{\ell}\right)$, where $J$ is the number of iterations of Algorithm \ref{alg:Max_SILNR}. We refer the details of the computational complexity analysis in \cite{choi2019joint}, in which the matrix inverse and the multiplication operations have shown to be performed in a divide and conquer manner by exploiting the block diagonal structure in ${\bf \bar A}_{\ell}\left({\bf f}_{\ell}\right)$ and ${\bf\bar B}_{\ell}\left({\bf f}_{\ell},\lambda\right)$.  For the convergence, the number of iterations for the inner loop is five at most in an average sense regarding random channels.
 
 \vspace{0.2cm} 
{\bf Remark 2 (The second-order optimality condition):} From numerical results, we observe that, in every case, the solution obtained from the proposed algorithm satisfies the second-order optimality condition derived in Theorem \ref{thm2}. We conjecture that the proposed algorithm ensures converging to a local-optimal solution.




\section{Simulation Results}
This section provides both the link-level and system-level simulation results to gauge the ergodic sum-spectral efficiency gains of the proposed SILNR maximization precoding compared to the existing precoding schemes. 
 

\begin{figure}[t]
\subfigure[]{
\includegraphics[width=\linewidth]{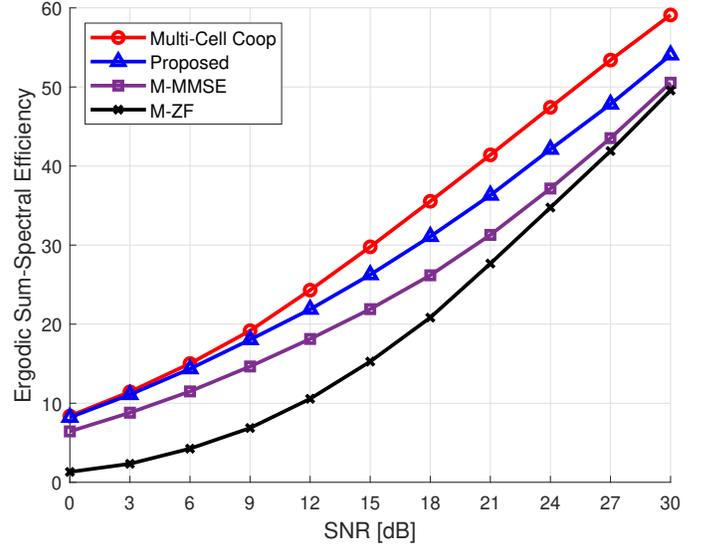}
\label{fig:LLS_Sumrate}
}
\subfigure[]{
\includegraphics[width=\linewidth]{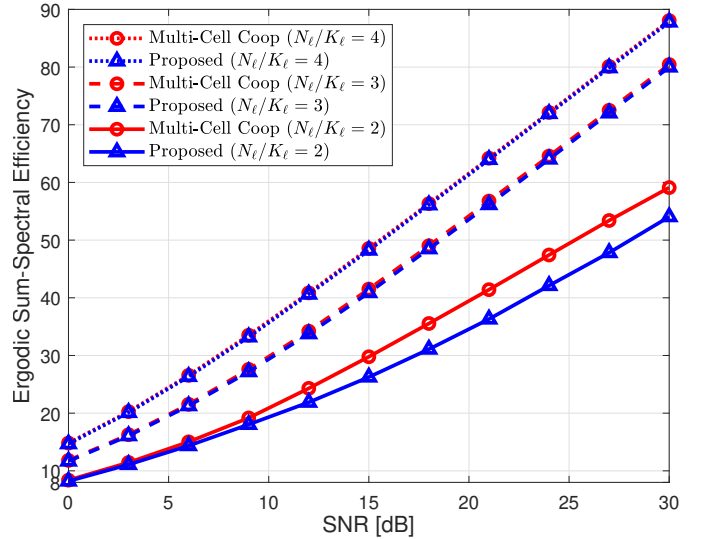}
\label{fig:IncreasingAntennas}
}
\caption{(a) The ergodic sum-spectral efficiencies for different precoding strategies and (b) the ergodic sum-spectral efficiencies when increasing the ratio of the number of antennas to the number of users without considering the zero-IUI condition.
}
\label{fig:LLS}
\end{figure}
\subsection{Link-Level Simulations}
We consider a two-cell scenario in which each BS equipped with $N_{\ell}$ $(=\!\!\!16)$ antennas serves single antenna eight users \cite{shin2010design}.    For link-level simulations, we assume that all channel vectors are drawn from ${\bf h}_{j,\ell,k}\sim\mathcal{CN}(0,{\bf I}_{
N_{\ell}})$, and each BS has perfect knowledge of local CSIT. In this setting, we compare the ergodic sum-spectral efficiency for different precoding strategies:
\begin{itemize}
    \item Multi-Cell Coop \cite{choi2018joint,choi2019joint}: This scheme is a cooperative precoding technique that maximizes the sum-spectral efficiency of users in the cooperative area. We use this precoding method as a benchmark for an upper bound of our proposed noncooperative precoding technique.
    \item Multi-Cell MMSE (or Multi-Cell SLNR)\cite{sadek2007leakage}: 
    \begin{align}
        {\underset{{\bf f}_{\ell,k}\in \mathbb{C}^{N_{\ell}}}{\text{arg~max}}}
        \frac{\left|{\bf \hat h}_{\ell,\ell,k}^{\sf H}{\bf f}_{\ell,k}\right|^2}{{\bf f}_{\ell,k}^{\sf H}\left(\sum_{(j,i)\in\mathcal{U}_{\ell}}{\bf \hat h}_{\ell,j,i}{\bf \hat h}_{\ell,j,i}^{\sf H}+\frac{{\tilde\sigma}_{\ell,k}^2}{P_{\ell}}{\bf I}_{N_{\ell}K_{\ell}}\right){\bf f}_{\ell,k}}.\label{eq:M-MMSE}
    \end{align}
    This scheme uses the downlink precoding solution in \eqref{eq:M-MMSE} by the uplink-downlink duality \cite{jose2011pilot, patcharamaneepakorn2012equivalence,bjornson2017massive}. 
    \item Multi-Cell ZF \cite{spencer2004zero}: This scheme eliminates both IUI and ICI.
 \end{itemize}
 
\begin {table}[t]
\caption {Complexity Order Analysis. } \vspace{-0.1cm}\label{tab:Complexity_Order} 
  	 \begin{center}
  {
  {\scriptsize
  \begin{tabular}{ l | c }
    \hline
        List & Computation complexity order\\ \hline\hline
        Multi-Cell MMSE & $\begin{array}{lcl}
        &\mathcal{O}\left((|\mathcal{U}_{\ell}|+|\mathcal{K}_{\ell}|)^2N_{\ell}\right.\\
        &\left.+(|\mathcal{U}_{\ell}|+|\mathcal{K}_{\ell}|)^3\right)
        \end{array}$ \\\hline
        Computation of ${\bf \bar A}_{\ell}({\bf f}_{\ell}^{\star})$ & $\mathcal{O}\left(N_{\ell}^2K_{\ell}\right)$ \\\hline
        Computation of ${\bf \bar B}_{\ell}({\bf f}_{\ell}^{\star},\lambda)$  & $\mathcal{O}\left(N_{\ell}^2K_{\ell}\right)$ \\\hline
        Computation of $\left[{\bf \bar B}_{\ell}({\bf f}_{\ell}^{\star},\lambda)\right]^{-1}$  & $\mathcal{O}\left(N_{\ell}^2K_{\ell}\right)$ \\\hline
        Computation of $\left[{\bf \bar B}_{\ell}({\bf f}_{\ell}^{\star},\lambda)\right]^{-1}{\bf\bar A}_{\ell}({\bf f}_{\ell}^{\star}){\bf f}_{\ell}^{\star}$  & $\mathcal{O}\left(N_{\ell}^2K_{\ell}\right)$ \\\hline
        Proposed algorithm &$\mathcal{O}\left(JN_{\ell}^2K_{\ell}\right)$ \\\hline\hline
  \end{tabular}}}
\end{center}\vspace{-0.3cm}
\end {table}

{\bf Validation of quasi-optimality:} As shown in Fig. 2-(b), the proposed algorithm tightly achieves the performance of the multi-cell cooperative precoding scheme. Our simulation result implies that our precoding solution makes the zero-IUI phenomenon occur, even if $N_{\ell}/K_{\ell}$ is not-so-large.

{\bf Ergodic sum-spectral efficiency performance:} Fig. 2-(a) shows how the ergodic sum-spectral efficiency changes with increasing SNRs. The proposed precoding yields considerable gains compared to the existing precoding methods in all SNRs. This performance improvement comes from better utilization of spatial degrees of freedom (DoF) to mitigate both IUI and ICI. Specifically, the proposed SILNR jointly finds a set of served users, precoding vectors, and power allocation by optimally balancing IUI and the geometric mean of interference leakage signals. Therefore, it allows each BS to exploit the spatial DoF in a more efficient way to increase the sum-spectral efficiency of the desired cell, while simultaneously reducing leakage interference power towards the other cells.

\begin {table}[t]
\caption {Parameters for System-Level Simulations.} \vspace{-0.1cm}\label{tab:Sys_Assumption} 
  	 \begin{center}
  	 {\footnotesize
  \begin{tabular}{ l  c }
    \hline\hline
    Parameters & Value  \\ \hline
        Topology of BS & 28 BSs over 7 hexagonal coverages \\ 
         \# of BSs; (MBS, PBS)  & (7,21)\\        
        Topology of user & Uniformly distributed per cell  \\ 
         \# of UEs per (MBS, PBS)  & (16,4)\\
         Bandwidth & 20 MHz \\
         Carrier frequency & 2 GHz \\
         MBS transmission power & 46 dBm \\
         PBS transmission power & 23 dBm \\
         Noise power & -113 dB\\
         Spatial channel model & Spatially correlated model\\
         Path-loss model & Okumaura-Hata urban model \\
         BS and user height & 32 m/1.5 m\\
         Channel estimation & MMSE estimator in \eqref{eq:Est_Err_Cov} \\ 
         Channel coherence time slot & ${\tau}_{\rm c} = 200$ \\
         Uplink training slot  & ${\tau}_{\rm u} = 28$ and ${\tau}_{\rm d}=0$\\
         Stopping condition & $\epsilon=0.1$\\\hline
  \end{tabular}}
\end{center}\vspace{-0.3cm}
\end {table}

{\bf Complexity order analysis:} We provided the computational complexity analysis of the proposed precoding method, which is summarized in Table \ref{tab:Complexity_Order}. The computational complexity of the proposed precoding method increases in the order of $\mathcal{O}\left(N_{\ell}^2K_{\ell}\right)$. To accomplish this analysis, we need to calculate the complexity order for computing functional matrices ${\bf\bar A}_{\ell}({\bf f}_{\ell}^{\star})\in\mathbb{C}^{N_{\ell}K_{\ell}\times N_{\ell}K_{\ell}}$ and ${\bf\bar B}_{\ell}({\bf f}_{\ell}^{\star},\lambda)\in\mathbb{C}^{N_{\ell}K_{\ell}\times N_{\ell}K_{\ell}}$. Exploiting the block diagonal structure of them, one can readily show that the complexity order becomes $\mathcal{O}\left(N_{\ell}^2K_{\ell}\right)$. In addition, using the fact that ${\bf\bar B}_{\ell}({\bf f}_{\ell}^{\star},\lambda)$ is the sum of rank-one matrices, one can compute the inverse matrix of it successively using Sherman-Morrison Lemma (or also known as Woodbury inverse) \cite{sherman1949adjustment}. Thanks to the block diagonal structure, computing the inverse of ${\bf\bar B}_{\ell}({\bf f}_{\ell}^{\star},\lambda)$ requires the computational complexity order of $\mathcal{O}\left(N_{\ell}^2K_{\ell}\right)$. By embracing the number of iterations to converge, $J$, the total computational complexity of the proposed precoding algorithm is $\mathcal{O}\left(JN_{\ell}^2K_{\ell}\right)$.

\subsection{System-Level Simulations}
\begin{figure}[t]
	\centering
    \includegraphics[width=\linewidth]{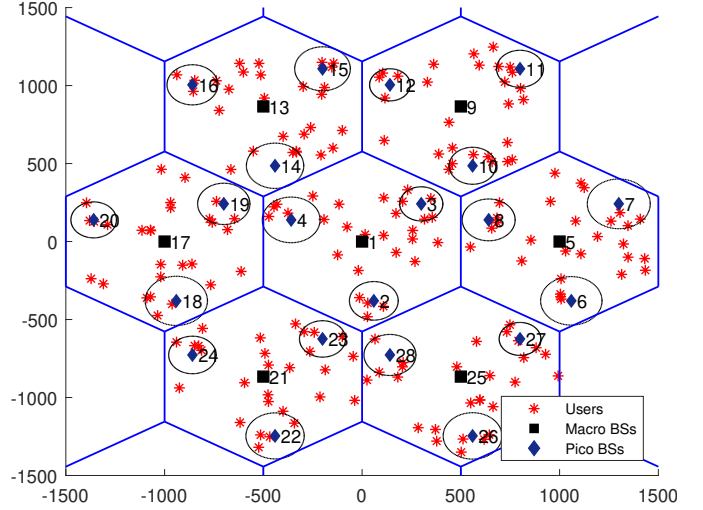}
  \caption{A snapshots of the network topologies used in system-level simulations. } \label{fig:Snapshot}
\end{figure}
Fig. \ref{fig:Snapshot} depicts the network topology for system-level simulations, which consists of seven hexagonal cells, each macro BS (MBS) coverage area contains three pico BSs (PBSs). See Table \ref{tab:Sys_Assumption} for the details of simulation parameters \cite{3GPP}.


For a fair comparison, we also consider the Single-Cell WMMSE \cite{christensen2008weighted} precoding method with a leakage level constraint. For this, we consider the following precoding optimization problem:
\begin{subequations}
\label{eq:WMMSE_Optimization_Problem}
\begin{align}
    &{\underset{\left\{{\bf f}_{\ell,k},w_{k},{ u}_k|k\in{\mathcal{K}}_{\ell}\right\}}{\text{arg~min}}} 
    \sum_{k=1}^{{K_{\ell}}}
    \left(
    w_ke_k-\log w_k
    \right),\label{eq:WMMSE_Obejective}\\
    &\text{subject to} ~~ \sum_{k=1}^{K_{\ell}} \|{\bf f}_{\ell,k}\|_2^2 \le 1,\label{eq:WMMSE_Power_Constraint}\\
    & ~~~~~~~~~~~~~~{\bf f}_{\ell,k}^{\sf H}\left(\sum_{(m,n)\in\mathcal{U}_{\ell}}{\bf h}_{\ell,m,n}{\bf h}_{\ell,m,n}^{\sf H}\right){\bf f}_{\ell,k} \le r_k,\label{eq:WMMSE_Leakage_Contraint}
\end{align}
\end{subequations}
where 
\begin{align}
    e_k&={ u}_k^{\sf H}\left(\sum_{i=1}^{K_{\ell}}\left|{\bf h}_{\ell,k}^{\sf H}{\bf f}_{\ell,i}\right|^2+\sigma^2\right){ u}_k-2{\sf Re}\left\{{ u}_{k}^{\sf H}{\bf h}_{\ell,k}^{\sf H}{\bf f}_{\ell,k}\right\}+1,\nonumber\\
    w_k&=e_k^{-1},\nonumber\\
    { u}_k&=\left(\sum_{i=1}^{K_{\ell}}\left|{\bf h}_{\ell,k}^{\sf H}{\bf f}_{\ell,i}\right|^2+\sigma^2\right)^{-1}{\bf h}_{\ell,k}^{\sf H}{\bf f}_{\ell,k},
\end{align}  denote MSE, MSE weight, and the optimal receiver, respectively. Also, $r_k$ dentoes a target leakage power level. The problem \eqref{eq:WMMSE_Optimization_Problem} is a combination of the conventional WMMSE optimization \eqref{eq:WMMSE_Obejective}, \eqref{eq:WMMSE_Power_Constraint} and leakage power constraints \eqref{eq:WMMSE_Leakage_Contraint}.

\begin{figure}[t]
\centering
\includegraphics[width=\linewidth]{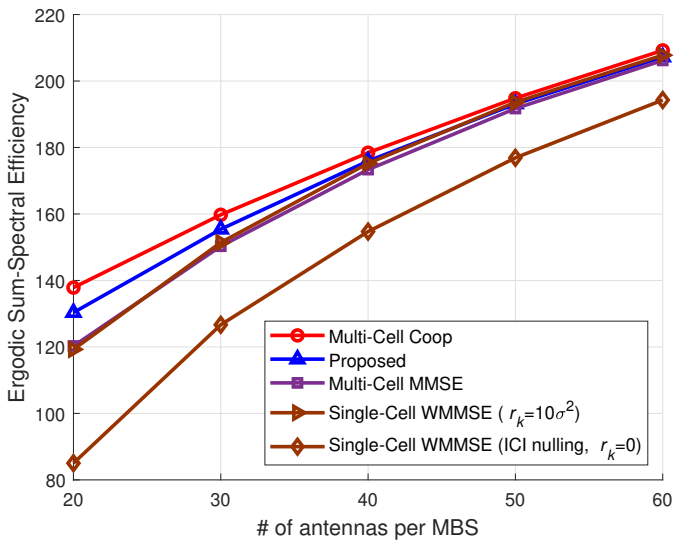}
\caption{The ergodic sum-spectral efficiencies for different precoding strategies with local and perfect CSIT.}
\label{fig:Perfect_Local_Antennas_16_4}
\end{figure}

{\bf Ergodic sum-spectral efficiency performance:} We evaluate how the ergodic sum-spectral efficiency behaves when increasing the number of antennas of MBSs. Fig. \ref{fig:Perfect_Local_Antennas_16_4} shows the ergodic sum-spectral efficiency performance under perfect local CSIT. As. can be seen, the proposed precoding method outperforms the existing precoding methods regardless of the number of antennas per MBS.  We capitalize that the existing noncooperative precoding can achieve the upper bound when the number of antennas is larger than $50$.

\begin{figure}[t]
\subfigure[]{
\includegraphics[width=\linewidth]{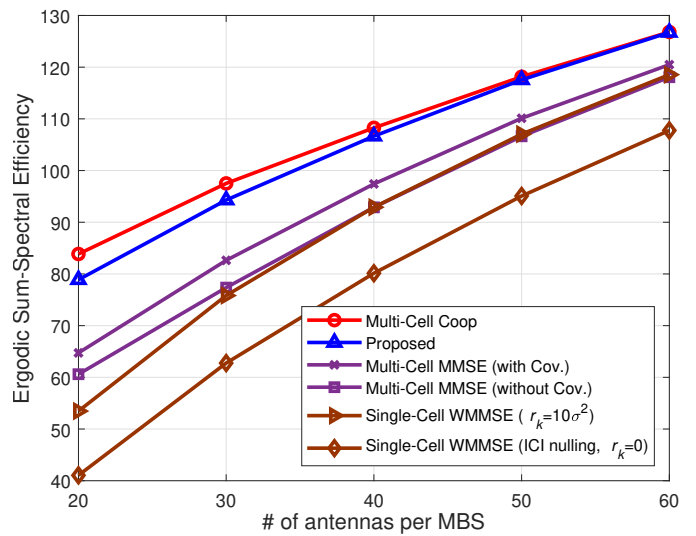}
\label{fig:Noisy_Local_Antennas_16_4}
}
\subfigure[]{
\includegraphics[width=\linewidth]{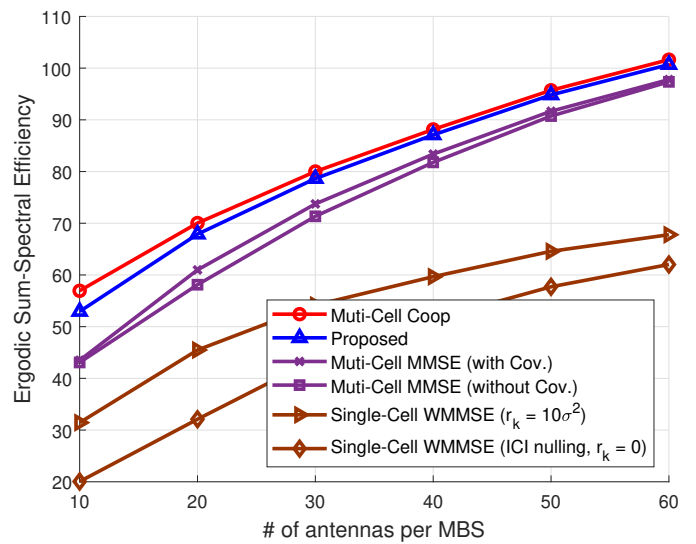}
\label{fig:Noisy_Local_Antennas_8_4}
}
\caption{The ergodic sum-spectral efficiencies for different precoding strategies with local and noisy CSIT when the number of UEs per (MBS,PBS) = (a) (16,4) and (b) (8,4).
}
\label{fig:Local_Antennas_16_4}
\end{figure}

 Fig. \ref{fig:Local_Antennas_16_4} demonstrates the ergodic sum-spectral efficiency performance under the noisy CSIT setting. For a fair comparison, we compare the performance of the proposed algorithm with multi-cell MMSE exploiting the covariance matrix ${\bf\Phi}_{\ell,\ell,k}$ \cite{li2015multi} (denoted by Muti-Cell MMSE (with Cov.)). As can be seen, the proposed algorithm obtains a substantial ergodic sum-spectral efficiency gain compared with other precoding methods in all antenna scales. This result implies that our proposed framework is robust to pilot contamination effects, a remarkable aspect of practical cellular systems.

\begin{figure}[t]
	\centering
    \includegraphics[width=\linewidth]{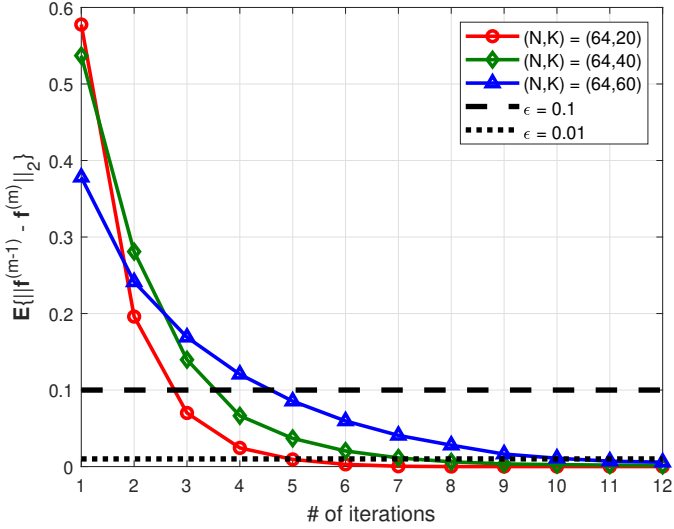}
  \caption{Convergence speed illustration with $\epsilon=0.1 \text{ and }\epsilon=0.01$.} \label{Fig_Convergence}
\end{figure}

{\bf Convergence speed of the proposed SILNR maximization precoding:} Fig. \ref{Fig_Convergence} illustrates the convergence speed of the proposed algorithm for the SILNR maximization precoding.  We consider the cases of $K=\{20,40, 60\}$ and $N=64$. We measure the mean square of the difference for the objective functions evaluated at two consecutive precoding solutions during iterations, i.e., $\mathbb{E}\left[\|{\bf f}_{\ell}^{(m)}-{\bf f}_{\ell}^{(m-1)}\|_2^2\right]$ where the average is taken over both the fading channel realizations and the user locations. As depicted in Fig. \ref{Fig_Convergence}, the number of required iterations to find the solution is at most five in an average sense, when we set the solution accuracy parameter to $\epsilon=0.1$. As improving the solution accuracy level to  $\epsilon=0.01$, ten iterations are sufficient to end the algorithm for all $K=\{20,40, 60\}$ and $N=64$. In addition, when the algorithm starts with the ZF precoding solution as an initial point, we empirically observe that the initially identified solution ${\bf f}^{\star}$ of Algorithm \ref{alg:Max_SILNR} has local optimality in the most of our simulations.

\section{Conclusion}
 
In this paper, we presented a novel distributed precoding technique using local CSIT for MU-MIMO HetNets. The central idea of the proposed precoding method was to maximize the downlink sum-spectral efficiency per cell while mitigating the other cell interference leakage using local CSIT. We introduced a new metric called SILNR that measures the ratio between the desired signal power and the superposition of IUI and interference leakage powers towards the other cells. Using this metric, we formulated a maximization problem of the product of SILNRs, which is a non-convex optimization problem. We derived the first- and the second-order necessary conditions for the local-optimal solution of this non-convex optimization problem. Leveraging these conditions, we presented a computationally efficient algorithm that ensures finding the local-optimal solution iteratively. Using both link-level and system-level simulations, we demonstrated our precoding method using local CSIT achieves the same sum-spectral efficiency of the cooperative precoding method that requires global CSIT when increasing the number of antennas BS, confirming that a synergetic gain is possible when massive MIMO meets HetNets.

 
 \appendix
 
 \subsection{Proof for Theorem 1}
 We commence by defining the Lagrange function:
\begin{align}
    \mathcal{L}\left({\bf f}_{\ell},\lambda\right) = \prod_{k=1}^{K_{\ell}}
    \frac{{\bf f}_{\ell}^{\sf H}\left({\bf\tilde A}_{\ell,\ell,k}+{\bf\hat L}_{\ell}\left({\bf f}_{\ell}\right)^{\frac{|\mathcal{U}_{\ell}|}{K_{\ell}}}\right){\bf f}_{\ell}}
    {{\bf f}_{\ell}^{\sf H}\left({\bf\tilde B}_{\ell,\ell,k}+{\bf\hat L}_{\ell}\left({\bf f}_{\ell}\right)^{\frac{|\mathcal{U}_{\ell}|}{K_{\ell}}}\right){\bf f}_{\ell}}
    -\lambda\left(\left\|{\bf f}_{\ell}\right\|_2^2-1\right).
\end{align}
To find a stationary point, we take the partial derivatives of $ \mathcal{L}({\bf f}_{\ell},\lambda)$ with respective to ${\bf f}_{\ell}^{\sf H}$ and $\lambda$, and set to them zero. When taking the derivative with respective to ${\bf f}_{\ell}^{\sf H}$, we obtain   
\begin{align}
    &\nabla_{{\bf f}^{\sf H}}\left\{\mathcal{L({\bf f}_{\ell},\lambda)}\right\}={\bf 0}\nonumber\\
    &\!\!\Leftrightarrow\!\!\sum_{i=1}^{K_{\ell}}
    \left(
    \prod_{k\ne i}{\bf f}_{\ell}^{\sf H}\left({\bf\tilde A}_{\ell,\ell,k}+{\bf\hat L}_{\ell}\left({\bf f}_{\ell}\right)^{\frac{|\mathcal{U}_{\ell}|}{K_{\ell}}}\right){\bf f}_{\ell}
    \right)\!\!
    \left(
    {\bf\tilde A}_{\ell,\ell,i}+\nabla_{{\bf f}_{\ell}^{\sf H}}{\bf\hat L}_{\ell}\left({\bf f}_{\ell}\right)^{\frac{|\mathcal{U}_{\ell}|}{K_{\ell}}}
    \right)\nonumber\\
    &\cdot\prod_{k=1}^{K_{\ell}}{\bf f}_{\ell}^{\sf H}\left({\bf\tilde B}_{\ell,\ell,k}+{\bf\hat L}_{\ell}\left({\bf f}_{\ell}\right)^{\frac{|\mathcal{U}_{\ell}|}{K_{\ell}}}\right){\bf f}_{\ell}\cdot{\bf f}_{\ell}\nonumber\\
    &\!\!=\sum_{i=1}^{K_{\ell}}
    \left(
    \prod_{k\ne i}{\bf f}_{\ell}^{\sf H}\left({\bf\tilde B}_{\ell,\ell,k}+{\bf\hat L}_{\ell}\left({\bf f}_{\ell}\right)^{\frac{|\mathcal{U}_{\ell}|}{K_{\ell}}}\right){\bf f}_{\ell}
    \right)\!\!
    \left(
    {\bf\tilde B}_{\ell,\ell,i}+\nabla_{{\bf f}_{\ell}^{\sf H}}{\bf\hat L}_{\ell}\left({\bf f}_{\ell}\right)^{\frac{|\mathcal{U}_{\ell}|}{K_{\ell}}}
    \right)\nonumber\\
    &\cdot\prod_{k=1}^{K_{\ell}}{\bf f}_{\ell}^{\sf H}\left({\bf\tilde A}_{\ell,\ell,k}+{\bf\hat L}_{\ell}\left({\bf f}_{\ell}\right)^{\frac{|\mathcal{U}_{\ell}|}{K_{\ell}}}\right){\bf f}_{\ell}\cdot{\bf f}_{\ell}-\lambda{\bf f}_{\ell}. \label{eq:derivative_f}
   \end{align} 
   By rearranging \eqref{eq:derivative_f}, we obtain:
\begin{align}    
    & \sum_{i=1}^{K_{\ell}}
    \left(
    \prod_{k\ne i}{\bf f}_{\ell}^{\sf H}\left({\bf\tilde A}_{\ell,\ell,k}+{\bf\hat L}_{\ell}\left({\bf f}_{\ell}\right)^{\frac{|\mathcal{U}_{\ell}|}{K_{\ell}}}\right){\bf f}_{\ell}
    \right)\!\!
    \left(
    {\bf\tilde A}_{\ell,\ell,i}+{\bf\tilde C}_{\ell,\ell,i}
    \right){\bf f}_{\ell}\nonumber\\
    &\!\!=\gamma({\bf f}_{\ell})\left[\sum_{i=1}^{K_{\ell}}
    \left(
    \prod_{k\ne i}{\bf f}_{\ell}^{\sf H}\left({\bf\tilde B}_{\ell,\ell,k}+{\bf\hat L}_{\ell}\left({\bf f}_{\ell}\right)^{\frac{|\mathcal{U}_{\ell}|}{K_{\ell}}}\right){\bf f}_{\ell}
    \right)\!\!
    \left(
    {\bf\tilde B}_{\ell,\ell,i}+{\bf\tilde C}_{\ell,\ell,i}
    \right)\right.\nonumber\\
&\left.-\frac{\lambda}{\gamma({\bf f}_{\ell})}\right]{\bf f}_{\ell}.
    \label{eq:proof_1}
\end{align}
The condition \eqref{eq:proof_1} simplifies to
 \begin{align}
    {\bf \bar A}_{\ell}\left({\bf f}_{\ell}\right){\bf f}_{\ell}=
    \gamma\left({\bf f}_{\ell}\right){\bf \bar B}_{\ell}\left({\bf f}_{\ell},\lambda\right){\bf f}_{\ell}.
\end{align}
We also take the partial derivatives of $\mathcal{L}({\bf f},\lambda)$ with respective to Lagrange multiplier $\lambda$ and set to them zero, which yields the condition:
\begin{align}
    \left\|{\bf f}_{\ell}\right\|_2^2-1=0 \Leftrightarrow \left\|{\bf f}_{\ell}\right\|_2^2=1.
\end{align}
This completes the proof.

\subsection{Proof for Theorem 2}
\begin{figure*}[t]
\begin{align}
\nabla_{{\bf f}^{\sf H}}^2\mathcal{L}\left({\bf f}_{\ell},\lambda\right)
   &=\left[
    \sum_{i=1}^{K_{\ell}}
    \frac{{\bf X}_{\ell,\ell,i}^{A}{\bf f}_{\ell}}
    {{\bf f}_{\ell}^{\sf H}{\bf Y}_{\ell,\ell,i}^{A}{\bf f}_{\ell}}-
    \frac{{\bf X}_{\ell,\ell,i}^{B}{\bf f}_{\ell}}
    {{\bf f}_{\ell}^{\sf H}{\bf Y}_B{\bf f}_{\ell}}
    \right]
    \cdot\left[
    \sum_{i=1}^{K_{\ell}}
    \frac{{\bf X}_{\ell,\ell,i}^{A}{\bf f}_{\ell}}
    {{\bf f}_{\ell}^{\sf H}{\bf Y}_{\ell,\ell,i}^{A}{\bf f}_{\ell}}-
    \frac{{\bf X}_{\ell,\ell,i}^{B}{\bf f}_{\ell}}
    {{\bf f}_{\ell}^{\sf H}{\bf Y}_B{\bf f}_{\ell}}
    \right]^{\sf H}
    +\gamma({\bf f}_{\ell})
    \left[
    \sum_{i=1}^{K_{\ell}}
    \frac{{\bf X}_{\ell,\ell,i}^{A}}{{\bf f}_{\ell}^{\sf H}{\bf Y}_{\ell,\ell,i}^A{\bf f}_{\ell}}
    +\frac{{\bf Z}_{\ell,\ell,i}}{{\bf f}_{\ell}^{\sf H}{\bf Y}_{\ell,\ell,i}^B{\bf f}_{\ell}}
    +\frac{{\bf X}_{\ell,\ell,i}^B{\bf f}_{\ell}{\bf f}_{\ell}^{\sf H}{\bf X}_{\ell,\ell,i}^B}{\left({\bf f}_{\ell}^{\sf H}{\bf Y}_{\ell,\ell,i}^B{\bf f}_{\ell}\right)^2}\right]\nonumber\\ 
    &-\left[\gamma({\bf f}_{\ell})
    \left\{\sum_{i=1}^{K_{\ell}}
    \frac{{\bf X}_{\ell,\ell,i}^{B}}{{\bf f}_{\ell}^{\sf H}{\bf Y}_{\ell,\ell,i}^B{\bf f}_{\ell}}
    +\frac{{\bf Z}_{\ell,\ell,i}}{{\bf f}_{\ell}^{\sf H}{\bf Y}_{\ell,\ell,i}^A{\bf f}_{\ell}}
    +\frac{{\bf X}_{\ell,\ell,i}^A{\bf f}_{\ell}{\bf f}_{\ell}^{\sf H}{\bf X}_{\ell,\ell,i}^A}{\left({\bf f}_{\ell}^{\sf H}{\bf Y}_{\ell,\ell,i}^A{\bf f}_{\ell}\right)^2}
    \right\}+\lambda{\bf I}\right].\label{eq:Hessian_Final2}
\end{align}
\hrulefill
\end{figure*}
 To prove the local optimality claim, it is sufficient to show that the extended Hessian matrix considering constraint sets at a stationary point is negative definite. To accomplish this, we derive the Hessian matrix evaluated at an arbitrary point ${\bf f}_{\ell}\in \mathbb{C}^{N_{\ell}K_{\ell}\times 1}$. For ease of exposition, we invoke the objective function $\gamma({\bf f})$ to the gradient of the Lagrange function as
\begin{align}
    &\nabla_{{\bf f}_{\ell}^{\sf H}}\left\{\mathcal{L}\left({\bf f}_{\ell},\lambda\right)\right\}
    =2\gamma({\bf f})
    \eta({\bf f}_{\ell})
    -
    2\lambda{\bf f}_{\ell}, \label{eq:Gradient}
\end{align}
where 
\begin{align}
	\eta({\bf f}_{\ell})= \left[
    \sum_{i=1}^{K_{\ell}}
    \frac{{\bf X}_{\ell,\ell,i}^{A}}
    {{\bf f}_{\ell}^{\sf H}{\bf Y}_{\ell,\ell,i}^{A}{\bf f}_{\ell}}
    -\frac{{\bf X}_{\ell,\ell,i}^{B}}
    {{\bf f}_{\ell}^{\sf H}{\bf Y}_{\ell,\ell,i}^{B}{\bf f}_{\ell}}
    \right]{\bf f}_{\ell}.
\end{align}
The extended Hessian matrix is obtained by directly calculating the gradient of \eqref{eq:Gradient} again, which is given by
\begin{align}
     \nabla_{{\bf f}_{\ell}^{\sf H}}^2\mathcal{L}\left({\bf f}_{\ell},\lambda\right)
     &=\nabla_{{\bf f}_{\ell}^{\sf H}}\left\{\nabla_{{\bf f}_{\ell}^{\sf H}}\left\{\mathcal{L}\left({\bf f}_{\ell},\lambda\right)\right\}\right\}\nonumber\\
     &=2\nabla_{{\bf f}_{\ell}^{\sf H}}\left\{\gamma({\bf f}_{\ell})\right\}\eta({\bf f}_{\ell})^{\sf H} 
     +2\gamma({\bf f}_{\ell})\nabla_{{\bf f}_{\ell}^{\sf H}}\left\{\eta({\bf f}_{\ell})\right\}-2\lambda{\bf I}.
    \label{eq:Hessian_matrix_1}
\end{align}

We compute $\nabla_{{\bf f}^{\sf H}}^2\left\{\mathcal{L}\left({\bf f}_{\ell},\lambda\right)\right\}
$ and $\nabla_{{\bf f}_{\ell}^{\sf H}}\left\{\eta({\bf f}_{\ell})\right\}$ in terms of the functional matrices defined as the Hermitian and positive definite matrices ${\bf X}_{\ell,\ell,i}^{A}$, ${\bf X}_{\ell,\ell,i}^{B}$, ${\bf Y}_{\ell,\ell,i}^{A}$, ${\bf Y}_{\ell,\ell,i}^{B}$, and ${\bf Z}_{\ell,\ell,i}$ in Theorem 2. Since the sum of positive definite (PSD) matrices is also a PSD matrix, the extended Hessian matrix becomes a negative definite matrix, provided that the the minimum eigenvalue of the third term in \eqref{eq:Hessian_Final2} is greater than the maximum eigenvalue of the sum of the first and second terms in \eqref{eq:Hessian_Final2}.  This is a sufficient condition for the local optimality. This ends the proof.

\bibliographystyle{IEEEtran}
\bibliography{IEEEabrv,SILNR_bib}

\end{document}